\documentclass[a4paper,11pt]{article}
\pdfoutput=1
\usepackage{jheppub}
\usepackage[T1]{fontenc}
\usepackage{epsfig}
\usepackage{subfigure}
\usepackage{float}
\usepackage{verbatim} 
\usepackage{pstricks}
\usepackage{color}
\usepackage{slashed}
\usepackage{multirow}
\usepackage{pstricks}
\usepackage{color}
\usepackage{booktabs}
\usepackage{subfigure}
\usepackage{epstopdf}
\usepackage[normalem]{ulem}

\def\tab#1{{Table~(\ref{#1})}}
\def\fig#1{{Figure~\ref{#1}}}

\def\eq#1{{Eq.~(\ref{#1})}}

\def\mg5{\textsc{MG5\_aMC}}
\def\Pythia8{\textsc{Pythia8}}

\begin{document}

\title{Scalar production and decay to top quarks including interference effects at NLO in QCD in an EFT approach }
\author[a]{Diogo Buarque Franzosi,}
\author[b]{Eleni Vryonidou,}
\author[c,d]{and Cen Zhang}

\abstract{ 
Scalar and pseudo-scalar resonances decaying to top quarks are common
predictions in several scenarios beyond the standard model (SM) and are
extensively searched for by LHC experiments. Challenges on the experimental
side require optimising the strategy based on accurate predictions.  Firstly,
QCD corrections are known to be large both for the SM QCD background and for
the pure signal scalar production. Secondly, leading order and approximate
next-to-leading order (NLO) calculations indicate that the interference between
signal and background is large and drastically changes the lineshape of the
signal, from a simple peak to a peak-dip structure.  Therefore, a robust
prediction of this interference at NLO accuracy in QCD is necessary to ensure
that higher-order corrections do not alter the lineshapes. We compute the exact
NLO corrections, assuming a point-like coupling between the scalar and the
gluons and consistently embedding the calculation in an effective field theory
within an automated framework, and present results for a representative set of
beyond the SM benchmarks. The results can be further matched to parton shower
simulation, providing more realistic predictions. We find that NLO corrections
are important and lead to a significant reduction of the uncertainties. We also
discuss how our computation can be used to improve the predictions for physics
scenarios where the gluon-scalar loop is resolved and the effective approach is
less applicable. } 
\affiliation[a]{II. Physikalisches Institut, Universit\"at G\"ottingen, Friedrich-Hund-Platz 1, 37077 G\"ottingen, Germany}
\affiliation[b]{Nikhef, Science Park 105, 1098 XG, Amsterdam, The Netherlands}
\affiliation[c]{Department of Physics, Brookhaven National Laboratory, Upton, NY 11973, USA}
\affiliation[d]{Institute of High Energy Physics, Chinese Academy of Sciences,
Beijing, 100049, China}

\emailAdd{dbuarqu@gwdg.de,eleniv@nikhef.nl,cenzhang@ihep.ac.cn}

\preprint{}


\maketitle

\section{Introduction}

The top quark is the heaviest of the standard model (SM) fermions, with a mass
close to the electroweak scale and a coupling to the Higgs boson close to
unity. These properties could indicate its intimate connection to the dynamics
of electroweak symmetry breaking (EWSB) and its role as a portal to physics
beyond the SM (BSM).  As an efficient top-quark factory, the LHC will inspect
the properties of this particle to an unprecedented level of precision, and set
accurate limits on any anomalous production mechanism.  The main production
mechanism of top quarks at the LHC is pair production initiated by gluons,
followed with roughly one third of the total cross section by single top
production. 

If a hypothetical new physics signal in top quark production originates from
scales out of the reach of the LHC, it will manifest as low energy
modifications parametrised by effective operators~\cite{AguilarSaavedra:2008zc,
Zhang:2010dr,
Degrande:2010kt,Choi:2012fc,Franzosi:2015osa,Zhang:2016omx,Englert:2016aei,Cirigliano:2016nyn}.
Alternatively, the LHC may be able to directly produce resonances which decay
into a pair of top quarks, producing a bump in a particular mass region.
Resonant top-pair production has been searched for by different experiments,
e.g.~\cite{Sirunyan:2017uhk,Aaltonen:2009tx,Khachatryan:2015sma,Chatrchyan:2013lca,Aad:2015fna}. 

Resonances decaying into a $t\bar{t}$ system exist in many BSM scenarios, such
as the two-Higgs-doublet model (2HDM), the supersymmetric models and several
models of dynamical EWSB.  Particularly relevant are the colour singlet
pseudo-scalar resonances (A) H decaying dominantly to tops, 
\begin{equation}
pp\to A/H \to t\bar{t}\,,
\label{eq:proc}
\end{equation}
due to the fact that in many models scalars have suppressed couplings to light
fermions, proportional to their masses. 
In order to correctly describe this process at the LHC, and to determine the
constraints on the parameters of different models, it is important to get
trustworthy and robust theoretical predictions. It is now known that QCD
corrections to this process have a large effect on the total cross section as
well as the differential distributions, both for the SM QCD production, known
at next-to-next-to-leading order in QCD and next-to-next-to-leading log soft
gluon resummation~\cite{Czakon:2013goa,Cacciari:2011hy} and NLO
EW~\cite{Bernreuther:2006vg,Czakon:2017wor}, and for (pseudo-)scalar
production, which is known at N$^3$LO in the infinite top mass limit
~\cite{Anastasiou:2015ema}) and NLO for finite top mass
\cite{Spira:1995rr,Harlander:2005rq}.

It is also well known that the interference between resonant signal and the QCD
background can drastically modify the lineshape of the signal, from a single
peak to a dip-peak structure, and may be even larger than the pure signal.
This effect has been studied at the leading-order (LO) in
Refs.~\cite{Gaemers:1984sj,Dicus:1994bm} and in more detail in
Refs.~\cite{Frederix:2007gi,Gori:2016zto,Jung:2015gta,Djouadi:2016ack}.  It
appears to receive similar QCD corrections, as indicated by estimates in an
effective field theory (EFT) approach in the soft gluon
approximation~\cite{Bernreuther:2015fts}, and in a simplified $K$-factor
derived from the geometric mean of the background and signal
$K$-factors~\cite{Hespel:2016qaf}. 

In this work for the first time we provide the full perturbative calculation of
the process $ pp(\to A/H) \to t\bar{t}$, at NLO in QCD, including the
interference, in the EFT approach, without any further approximation. The
effective description is justified for BSM models in which new heavy QCD
charged states provide the dominant contribution in generating gluon-scalar
interactions.  Our calculation does not directly apply to the cases in which
the loop contribution is dominated by the top quark.  However, we show that in
such cases our calculation can be improved by adopting a reweighting technique,
in which the Born pieces of the calculation are described by the full one-loop
amplitudes~\cite{Frederix:2014hta,Maltoni:2014eza}.  In absence of a full
calculation which is currently out of reach, this approach is expected to
provide a reliable estimate for the QCD corrections even in the kinematic
region where the EFT approximation is not valid. Unlike
Ref.~\cite{Bernreuther:2015fts}, our reweighting includes the interference part
and is done on an event-by-event basis, and can be passed to the parton shower
(PS) simulation, thus providing more realistic predictions.

Our calculation is performed within the \textsc{MadGraph5\_aMC@NLO}
(\mg5)~\cite{Alwall:2014hca} framework which allows us to perform automatic
simulation of events with PS matching.  The computation of NLO QCD corrections
to top-quark processes involving anomalous couplings or higher-dimension
operators has evolved in recent
years~\cite{Drobnak:2010by,Drobnak:2010wh,Zhang:2013xya,Drobnak:2010ej,Zhang:2014rja,Liu:2005dp,Gao:2009rf,Zhang:2011gh,Li:2011ek,Wang:2012gp,Shao:2011wa,
Degrande:2014tta,Rontsch:2014cca,Franzosi:2015osa,Rontsch:2015una,Zhang:2016omx,Bylund:2016phk,Maltoni:2016yxb}.
In particular automated calculations within \mg5 have been performed in
\cite{Degrande:2014tta,Franzosi:2015osa,Zhang:2016omx,Bylund:2016phk,Maltoni:2016yxb},
and the implementation used here closely relates to these works. In the context
of the EFT, one needs to consistently take into account the running and mixing
between different operators, which could lead to additional complications.  In
the process being studied, an interesting feature that has not been discussed
before is that the gluon-scalar operator mixes into the chromo-magnetic
dipole-moment operator, and therefore both operators must be taken into account
in a consistent NLO calculation and put together in a coherent setup. Thanks to
our previous work in Ref.~\cite{Franzosi:2015osa}, the QCD corrections to the
chromo-magnetic operator are already available.  In this work, we will then
deal with the remaining problems, i.e.~the implementation of the operator
mixing, as well as the two-loop matching which is required to determine the
size of the chromo-magnetic operator.  Such a calculation is an interesting one
in itself: unlike the previous NLO EFT calculations in
Refs.~\cite{Drobnak:2010by,Drobnak:2010wh,Zhang:2013xya,Drobnak:2010ej,Zhang:2014rja,Liu:2005dp,Gao:2009rf,Zhang:2011gh,Li:2011ek,Wang:2012gp,Shao:2011wa,
Degrande:2014tta,Rontsch:2014cca,Franzosi:2015osa,Rontsch:2015una,Zhang:2016omx,Bylund:2016phk,Maltoni:2016yxb},
in this work the EFT is used in a top-down way, by starting from explicit BSM
models, performing matching calculations at the same accuracy, employing RG
equations down to the scale of the problem, and physical results are obtained
by carrying out an NLO calculation there.

The paper is organised as follows. A general description of the features and
strategy of the calculation is presented in Section \ref{sec:lineshape}. In
Section \ref{sec:theory} the theoretical setup is discussed in detail.
In Section \ref{sec:benchmarks} we provide the description of several
benchmark scenarios which are of phenomenological relevance. In Section
\ref{sec:results} we show the corresponding results for each benchmark and
discuss the effects of NLO corrections on the lineshape.  Finally we present
our conclusions in Section \ref{sec:conclusion}.

\section{Heavy scalar lineshape}
\label{sec:lineshape}

In this section we briefly discuss some basic features of the heavy scalar
lineshape in the $t\bar t$ final state.  In several BSM scenarios, the
additional scalar couplings to fermions are hierarchical, roughly proportional
to the fermion masses.  This is the reason why we focus on the gluon fusion
process as the dominant production mode.  Contributions from $q\bar q$ initial
states are relatively easier to deal with, and the interference between signal
and background vanishes at LO for a colour singlet resonance, and so we will
not consider them here.

The gluon-gluon-scalar vertex can take two forms depending on the
underlying model.  If the vertex is dominated by heavy particle loops
(e.g.~vector-like fermions or scalar top partners), or induced by strong
dynamics at a high scale, this interaction will not be resolved at the scale of
the scalar resonance.  In this case the vertex can be simply represented by the
dimension-five operators 
\begin{flalign}
	O_{HG}&=g_s^2G_{\mu\nu}^AG^{A\mu\nu}H\,\label{eq:ohg},
	\\
	O_{A\widetilde G}&=g_s^2G_{\mu\nu}^A\widetilde G^{A\mu\nu}A\,,
	\label{eq:oAg}
\end{flalign}
with $G_{\mu\nu}^A$ being the gluon field strength tensor and $\widetilde G^{A\mu\nu}$ its dual.  
On the other hand, if the vertex is loop-induced with lighter
particles, such as top quarks running in the loop, it will be resolved at the
resonance scale and the interaction will give rise to an absorptive phase.  It
is convenient to distinguish between these two cases, as they often lead to
different lineshapes, and the resolved case is more difficult to compute at NLO
accuracy.  In practice, one may need to deal with a mixed scenario, if there
are contributions from both light and heavy loop particles.

\subsection{Interference between signal and background}
\begin{figure}[htb]
	\begin{center}
	\begin{tabular}{c}
	    \includegraphics[width=.9\linewidth]{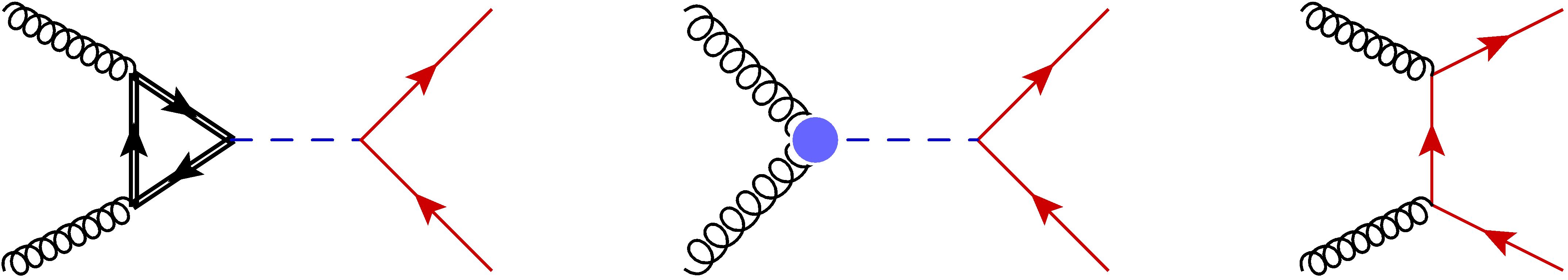}	\\
		\hspace{1.2cm}(a) \hspace{5cm}(b)\hspace{4.cm} (c)
	\end{tabular}		
	\end{center}
	\caption{LO signal and background.  Only one diagram from SM background is shown.}
	\label{fig:sblo}
\end{figure}

It has been noticed in an earlier work \cite{Dicus:1994bm}, and discussed in a
series of recent works \cite{Barcelo:2010bm,Barger:2011pu,Bai:2014fkl,
Craig:2015jba,Jung:2015gta,Djouadi:2016ack,Hespel:2016qaf,Carena:2016npr}, that
the production of a heavy scalar resonance leads to large interference with the
SM $t\bar t$ background.  This large interference could be further augmented by
a nontrivial relative phase between the signal and the SM background amplitude,
possibly leading to more complicated structures.  Possible lineshapes can vary
from a pure Breit-Wigner (BW) resonance to peak-dip structures, pure dip
structures, and even enhanced-peak structures, depending on the details of the
underlying model \cite{Jung:2015gta}.

To briefly explain these effects, in Figure~\ref{fig:sblo} we show the loop
induced resonant Feynman diagram (a), which in the heavy fermion limit can be
described by a contact interaction as in diagram (b), and a SM QCD background
diagram (c).  Due to the large production rate in the SM, the interference is
expected to be important. In particular for non-narrow resonances, the
interference can be larger than the signal.  

The impact of the interference on the lineshape can be understood by
considering the heavy scalar propagator convoluted with the loop form factor of
the top loop (we consider only the top quark loop as a resolved loop)  and the
contact interaction from high scale physics, which gives
\begin{equation}
{\cal M}_{sig} \propto \frac{c_g+c_t\,A^{H,A}_{1/2}\left(\frac{s}{4m_t^2}\right)}{s-M^2+iM\Gamma}\,,
\label{eq:amplo}
\end{equation}
with
\begin{eqnarray}
A^{A}_{1/2}(\tau)&=&f(\tau)/\tau  \,,\\
A^{H}_{1/2}(\tau)&=&\frac{3}{2\tau^2}\left(\tau+(\tau-1)f(\tau)\right)\,, \\
f(\tau) &=& 
    \begin{cases}
      \text{arcsin}^2(\sqrt{\tau}), & \quad \tau\leq 1\,, \\
      -\frac{1}{4}\left[\log\left(\frac{1+\sqrt{1-\tau^{-1}}}{1-\sqrt{1-\tau^{-1}}} \right)-i\pi \right]^2, & \quad \tau>1\,,
    \end{cases}
\end{eqnarray}
where $c_g$ is the effective contact interaction coupling and $c_t$ is the
normalised fermion coupling to the (pseudo-)scalar. The $A^{H,A}_{1/2}$ loop
form factors for scalar (H) and pseudo-scalar (A) approach $A^{H,A}_{1/2}\to 1$
in the limit of heavy fermion mass in the loop ($\tau\to 0$) and develop an
imaginary part for a resolved loop, $\tau>1$. 

The signal contribution comes from squaring the resonant amplitude and displays
a BW shape in the $m(t\bar t)$ spectrum, with small perturbations from the loop
form factor. The interference, on the other hand, is proportional to the real
part of ${\cal M}_{sig}$,
\begin{equation}
{\rm Re}{\cal M}_{sig} \propto \frac{(c_g+c_t{\rm Re}A^{S,A}_{1/2}(s/(4m_t^2)))(s-M^2)+{\rm Im}A^{S,A}_{1/2}(s/(4m_t^2))M\Gamma}{(s-M^2)^2+(M\Gamma)^2}\,.
\label{eq:rem}
\end{equation}
In the non-resolved case, the only complex phase in this problem comes from the
phase of the propagator, and thus the interference contribution depends only on
the real part of the propagator, which flips sign near the resonance due to the
factor of $s-M^2$ in its numerator.  Even more interesting is the resolved
case, where an additional complex phase will be provided by the loop form
factors.  In this case the second term in the numerator of the r.h.s of
\eq{eq:rem} contributes with a pure BW-shaped component to the full signal.
For example, in the 2HDM, the heavy scalar and pseudoscalar couple to the
gluons through top-quark loops.  In this case the phase starts to appear as the
mass of the resonance goes above the $2m_t$ threshold, possibly leading to pure
dip or enhanced peak structures \cite{Jung:2015gta}.

\subsection{NLO approach}

Searching for heavy scalars in the $t\bar t$ final state is challenging, not
only because of the complicated lineshapes, but also due to the top-quark
invariant mass reconstruction that smears the signal, and the systematic
uncertainty associated with the large production cross section of the SM
top-quark pairs \cite{Craig:2015jba,Carena:2016npr}.  For this reason, it is
useful to optimise the experimental search strategy according to theoretical
predictions for the lineshape, which have to include the interference effect
that determines the signal shape.  In fact, in a recent ATLAS search
\cite{ATLAS:2016pyq}, the interference indeed plays an important role in the
interpretation of the results in the 2HDM. On the theory side, however, the
interference contribution to the signal is currently known only at the LO
accuracy in QCD, while NLO corrections are expected to be important, given that
the corresponding corrections to the pure signal and background are both large
(at roughly $\sim 100\%$ and $\sim 50\%$).  Recent progress has been made in
Ref.~\cite{Bernreuther:2015fts}, where an EFT approximation together with a
soft-gluon approximation has been adopted.  More recently in
Ref.~\cite{Hespel:2016qaf}, predictions based on $K$-factors inferred from the
pure signal and background components have also been provided.  

In this work, we aim to follow a more solid approach to the NLO computation of
the interference, based on the EFT framework, which in the unresolved case
provides accurate results (i.e.~without using soft-gluon approximation), and in
the resolved case, can be further improved by using reweighting techniques (by
Born-reweighting or FTapprox which includes the exact real corrections
\cite{Frederix:2014hta,Maltoni:2014eza}) as we will discuss in the following.
These results will then be passed to PS simulation, to obtain more realistic
predictions.

\begin{figure}[htb]
	\begin{center}
		\includegraphics[width=.8\linewidth]{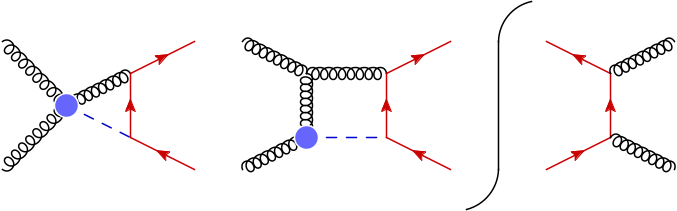}
	\end{center}
	\caption{Non-factorisable corrections (selected) to the interference at NLO.}
	\label{fig:nonfac}
\end{figure}
\begin{figure}[htb]
	\begin{center}
		\includegraphics[width=.66\linewidth]{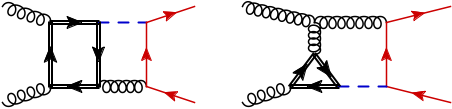}
	\end{center}
	\caption{Two-loop contributions involved in non-factorisable corrections.
	Loop particles depend on the details of the model.}
	\label{fig:2loop}
\end{figure}

Let us first discuss the main difficulty at NLO.  The signal is loop-induced
(by light or heavy particles), therefore computing the signal at NLO requires
calculating two-loop diagrams. For the signal, the initial state factorisable
corrections are well known \cite{Spira:1995rr,Harlander:2005rq} and also
implemented in Monte Carlo generators \cite{Mantler:2015vba}. Similarly the
final state corrections are also well known as part of the QCD corrections to
the Higgs decay width to heavy quarks \cite{Djouadi:1995gt}.  Non-factorisable
corrections vanish for the signal, because the non-factorisable virtual
corrections produce the top-quark pair in a colour octet state, and therefore
do not interfere with the Born level signal diagram which contains a colour
singlet.  Non-factorizable real corrections vanish for the same reason.  This
is however not true anymore when interference is taken into account, see
Figure~\ref{fig:nonfac}, where the interactions represented by blue dots can be
either resolved or non-resolved.  In this case virtual corrections that are
non-factorisable could interfere with the SM QCD background, giving an NLO
correction to the interference component.  If the $ggH(A)$ vertex is induced by
loop particles, a complete calculation of these corrections will involve
two-loop diagrams, e.g.~as shown in Figure~\ref{fig:2loop}.  These computations
are difficult to perform due to the many scales involved.

However, the computation of the interference in the unresolved case, mediated
by a gluon Higgs effective operator, can be performed exactly at NLO, and will
be presented for the first time in this work without further approximations.
As we will show in the next section, including only the  scalar-gluon operators
(\eq{eq:ohg}-\eq{eq:oAg}) is not enough for a consistent calculation in the
EFT.  This is because operator mixing into a top-quark dipole moment becomes
relevant in this process, and we will have to include this operator in our
approach.  This procedure also involves the matching stage, where a two-loop
matching needs to be performed also for the dipole operator.

We note here that the exact results in the EFT can be also used to improve the
predictions for the resolved case.  It is well known that the EFT approach does
not hold if $s$ goes above $4m^2_X$, where $X$ is the particle running in the
loop, but reweighting with the exact Born-level result helps to improve the
predictions for heavy scalar single or double production.  In particular it has
been shown that the EFT approximation with a Born reweighting is a good
approximation for single Higgs production even for heavy scalar mass
\cite{Kramer:1996iq} and thus we follow a similar approach to improve our
predictions for the resolved case.  While this is appropriate for the signal,
for the interference a simple reweighting based on the ratio between LO exact
and EFT results is not reliable, since reweighting is rather problematic as the
ratio $I_{\textrm{exact}}/I_{\textrm{EFT}}$ diverges as the exact and EFT
amplitudes cross zero at different points. In order to address this problem a
phase reproducing the exact LO behaviour can be introduced to the EFT
amplitude, by adding an ad-hoc imaginary part to the effective operator
coefficient. 

\section{Theoretical setup}
\label{sec:theory}
In this section we discuss the technical setup of our approach in more detail.

\subsection{EFT for the unresolved case}
We first discuss the case where the $ggH(A)$ vertex is mediated by high-scale
physics.  In this case the EFT is a good approximation.  Consider the following
EFT of the SM augmented by a heavy scalar $H$
\begin{flalign}
	L_{Eff}=L_{SM}+y_t\bar{t}tH+\frac{C_{HG}}{\Lambda}O_{HG},
	\label{eq:eff01}
\end{flalign}
where $O_{HG}$ is given in \eq{eq:ohg}.

\begin{figure}[htb]
	\begin{center}
		\includegraphics[width=.56\linewidth]{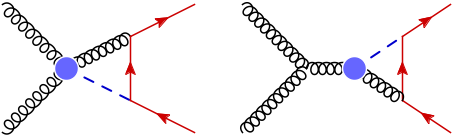}
	\end{center}
	\caption{Selected UV divergent diagrams in non-factorisable contributions from
		the Lagrangian in Eq.~(\ref{eq:eff01}).}
		\label{fig:uvpoles}
\end{figure}

If the full theory is known, one can determine the coefficient
by explicitly computing  the vertex in the full theory. For example, if this
contribution is induced by a vector-like fermion, $F$, via the following Yukawa
interaction:
\begin{flalign}
	L_{Yuk}=y_F\bar{F}FH\;,
	\label{eq:yukawa}
\end{flalign}
\begin{figure}[htb]
	\begin{center}
		\includegraphics[width=.56\linewidth]{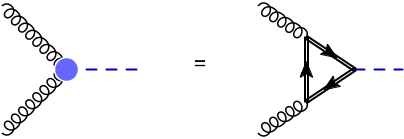}
	\end{center}
	\caption{One-loop matching to compute $C_{HG}$.}
	\label{fig:mathcing1}
\end{figure}
by computing diagrams shown in Figure~\ref{fig:mathcing1} one finds
\begin{flalign}
	\frac{C_{HG}(\mu)}{\Lambda}=-\frac{y_F}{48\pi^2m_F}+\mathcal{O}(\alpha_s)\,.
	\label{eq:1loopmatching}
\end{flalign}
By using this operator one can compute the process at NLO in QCD, which
involves at most one-loop diagrams.  For consistency, corrections to
Eq.~(\ref{eq:1loopmatching}) from two-loop matching should be incorporated
\cite{Spira:1995rr}:
\begin{flalign}
	\frac{C_{HG}(\mu)}{\Lambda}=-\frac{y_F}{48\pi^2m_F}-\frac{11\alpha_sy_F}{192\pi^3m_F}+\mathcal{O}(\alpha_s^2).
	\label{eq:2loopmatching}
\end{flalign}
However, this computation leads to uncancelled UV poles from non-factorisable
contributions, whenever a top quark, a gluon and a scalar form a loop, as shown
in Figure~\ref{fig:uvpoles}.

The reason for the UV divergence is not difficult to understand.  The effective
Lagrangian, given in Eq.~(\ref{eq:eff01}), is in principle not a renormalizable
one.  It is well known that the EFT is renormalisable only if one considers the
complete set of higher-dimensional operators up to a certain dimension.
However, since so far only the operator $O_{HG}$ is added, it could mix into a
different operator not included in the Lagrangian, leading to
non-renormalisability of the theory.  In fact, through the triangles shown in
Figure~\ref{fig:uvpoles}, the $O_{HG}$ induces a chromo-magnetic dipole
operator of the top quark, 
\begin{flalign}
	O_{tG}=g_sy_t\bar t\sigma^{\mu\nu}T^AtG^A_{\mu\nu}.
	\label{eq:otg}
\end{flalign}
The physical interpretation is the following.  The right-hand side of the
second diagram in Figure~\ref{fig:uvpoles} is a triangle loop made by a Higgs, a
top, and a gluon.  In the full theory, it represents a similar diagram but with
the blue dot replaced by a fermion loop, namely the Barr-Zee diagram
\cite{Barr:1990vd}, an important contribution for lepton dipole moments from
additional scalars.  We see here that the same diagram is playing a role also
for the top quark, leading to a chromo-magnetic dipole moment of the top.

Two comments are in order.  First, the $gg\to h$ calculation in the SM is based
on essentially the same Lagrangian, but the renormalisability is not an issue
there, because $O_{tG}$ simply does not enter the $gg\to h$ process. Here we see
that the problem occurs because we have a scalar that decays into the $t\bar t$
final state, which makes $O_{tG}$ relevant.  This also explains why the problem
occurs only in the non-factorisable pieces.  Second, one option to avoid this
problem is to simply assume that the gluons in these loops are always soft for the
dominant contribution, which is the assumption used in
Ref.~\cite{Bernreuther:2015fts}.  The assumption is a good one for the resonant
region, but its validity for predicting the lineshape in a much larger range is not
guaranteed. The process being considered here is most interesting if the
interference effect is large, which implies a large width and a non-trivial
lineshape over a much larger range of the $m(t\bar t)$ distribution. For this
reason, in this work we will not use this approximation.

It is clear that the solution is to take Eq.~(\ref{eq:eff01}) and extend it to incorporate
also the $O_{tG}$ operator
\begin{flalign}
	L_{Eff}=L_{SM}+y_t\bar{t}tH+\frac{C_{HG}}{\Lambda}O_{HG}+\frac{C_{tG}}{\Lambda}O_{tG}\,,
	\label{eq:eff02}
\end{flalign}
and perform the calculation. With the correct mixing coefficient,
\begin{flalign}
	&C_{tG}\rightarrow C_{tG}^{(0)}=Z_{tG,i}C_i
	\\
	&\delta Z_{tG,HG}=-\frac{\alpha_s}{2\pi}\epsilon_{UV}^{-1}
\end{flalign}
proper counterterms will be generated to cancel the UV poles of the diagrams in
Figure~\ref{fig:uvpoles}.

The missing piece now is to determine the coefficient of $C_{tG}$ in the EFT
via a matching procedure.   Without the matching, the contribution from
Figure~\ref{fig:uvpoles} will depend on the scheme in which we determine the
counterterms.

\subsubsection{Matching}

\begin{figure}[htb]
	\begin{center}
		\includegraphics[width=.86\linewidth]{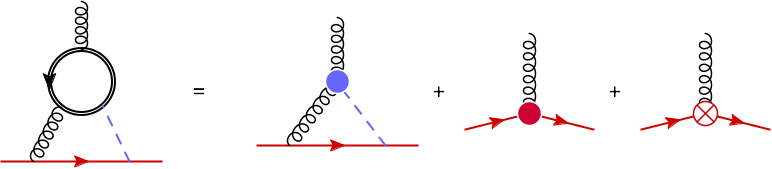}
	\end{center}
	\caption{Two-loop matching to determine the coefficient $C_{tG}$ at two-loop accuracy.}
		\label{fig:bzmatching}
\end{figure}

At the one loop level there is no contribution to $C_{tG}$.  However to be
consistent with the NLO calculation, we need to consider the matching at two
loops, just like what we have done for $C_{HG}$ in Eq.~(\ref{eq:2loopmatching}).
This can be done by computing the Barr-Zee contribution to the $gt\bar t$ function,
and matching the result to the EFT contribution. The actual computation can be
model dependent.  For a model with a CP-even scalar $H$ coupled to vector-like
fermions $F$ through a Yukawa coupling given by Eq.~(\ref{eq:yukawa}), this is
illustrated in Figure~\ref{fig:bzmatching}.
On the left-hand side, the full result for the Barr-Zee diagrams has been
computed in the context of lepton dipole moments.  We take the analytical
expression from Ref.~\cite{Altmannshofer:2015qra} and expand in powers of
$1/m_F$.  On the right-hand side, the one-loop contribution from $O_{HG}$ is
computed using $C_{HG}$ which has been matched at one loop.  The counterterm
from $C_{tG}$ is determined in the $\overline{MS}$ scheme.  Setting
$\Lambda=m_F$, and neglecting the top-quark mass, the matching is represented
by the following equation:
\begin{flalign}
	&-\frac{g_s^3y_Fy_t}{2304\pi^4m_F}\left( 12\log\frac{m_F}{m_H}+13 \right)
\nonumber\\
	&=-\frac{g_s^3y_Fy_t}{768\pi^4m_F}\left(\frac{2}{\epsilon}+ 4\log\frac{\mu}{m_H}+3 \right)
	+C_{tG}\frac{y_tg_s}{m_F}
	+\frac{g_s^3y_Fy_t}{768\pi^4m_F}\left(\frac{2}{\epsilon} \right).
	\label{eq:bzmatching}
\end{flalign}
The left-hand side is finite as expected from the full theory, and involves two
scales, $m_F$ and $m_H$.  The same $\log m_H$ dependence appears also on the
right-hand side, with exactly the same coefficient.  This is expected because
the two theories are supposed to describe the same IR physics.  On the other
hand, the loop contribution from the right-hand side is divergent, also
expected because the EFT modifies the UV structure of the theory.  By choosing
$\mu=m_F$, we can cancel the log terms in the matching, and the resulting
coefficients are defined at the scale $m_F$.  The UV pole and the remaining
finite terms from the loop contribution on the right have only tree-level
structure.  The former will be cancelled by the last term, i.e.~an
$\overline{MS}$ counterterm, while the latter will match the full result on the
left-handed side by choosing a proper value for $C_{tG}$.  We find:
\begin{flalign}
	C_{tG}(m_F)=-\frac{\alpha_s}{144\pi^3}y_F+\mathcal{O}(\alpha_s^2)\,.
	\label{eq:2looptg}
\end{flalign}
This together with Eq.~(\ref{eq:2loopmatching}) defines our EFT at the two-loop
accuracy.  Scheme dependence always cancel in the final results as long as the
matching and the actual calculation based on EFT are performed within the same
scheme.  This is because the sum of the last two terms in
Eq.~(\ref{eq:bzmatching}) is scheme independent.

The top-quark mass, $m_t$, has been neglected in the matching.  This will only
give rise to power corrections such as $(m_t/m_H)^2$.  Given the smallness of
$C_{tG}$, this contribution is expected to be negligible.

Adding a CP-odd scalar, $A$, will not affect the above approach.  Consider
extending Eq.~(\ref{eq:yukawa}) by
\begin{flalign}
	L_{Yuk}=y_F\bar{F}FH+\tilde y_{F}Fi\gamma^5FA\;,
	\label{eq:yukawa5}
\end{flalign}
where $A$ is a CP-odd scalar.  The theory is still CP-conserving, and after
matching will lead to our complete EFT:
\begin{flalign}
	L_{Eff}=L_{SM}+y_t\bar{t}tH+\tilde y_t\bar{t}i\gamma^5tA+\frac{C_{HG}}{\Lambda}O_{HG}
+\frac{C_{A\tilde G}}{\Lambda}O_{A\tilde G}+\frac{C_{tG}}{\Lambda}O_{tG},
	\label{eq:eff05}
\end{flalign}
with operators written explicitly in \eq{eq:ohg}, \eq{eq:oAg} and \eq{eq:otg}.
Note that $C_{tG}$ is real as long as CP is conserved.  The matching can be
performed in a similar way since the two-loop corrections for both $ggA$ and $\bar
ttg$ (from a CP-odd scalar) are known \cite{Spira:1995rr,Abe:2013qla}.  We find
\begin{flalign}
	&\frac{C_{HG}(m_F)}{\Lambda}=-\frac{y_F}{48\pi^2m_F}-\frac{11\alpha_sy_F}{192\pi^3m_F}+\mathcal{O}(\alpha_s^2) \,,
	\\
	&\frac{C_{A\tilde G}(m_F)}{\Lambda}=-\frac{\tilde y_F}{32\pi^2m_F}+\mathcal{O}(\alpha_s^2)\,,
	\\
	&\frac{C_{tG}(m_F)}{\Lambda}=-\frac{\alpha_s}{1152\pi^3m_F}\left( 8y_F-9\tilde y_F \frac{\tilde y_t}{y_t} \right)+\mathcal{O}(\alpha_s^2).
\end{flalign}
It is not our purpose to present matching results for all BSM extensions with
additional scalars.  However, the procedure outlined above can be carried out
for any perturbative BSM model.

\subsubsection{Running}

The actual calculation and simulation will be performed at the renormalisation
scale $\mu$, conventionally chosen at $\mu=m_H/2\ (m_A/2)$. In case the scale
$m_F$ and $\mu$ are well separated, one needs to run the EFT from $m_F$ down to
$\mu$ to resum the large logarithmic contributions.  For
$\mathcal{O}(\alpha_s)$ mixing this is done by solving the RG equations
\begin{flalign}
	\frac{dC_i(\mu)}{d\log\mu}=\frac{\alpha_s(\mu)}{\pi}\gamma_{ij}C_j(\mu)\,,
\end{flalign}
where for the three operator coefficients $C_{HG},C_{A\tilde G},C_{tG}$, the
matrix $\gamma_{ij}$ is given by
\begin{flalign}
	\gamma=\left(\begin{array}{ccc}
		0 & 0 & 0\\
		0 & 0 & 0\\
		-1 & \tilde y_t/y_t & 1/3
	\end{array}\right)\,.
\end{flalign}
The coefficients at a given scale $\mu$ are thus given by
\begin{flalign}
	C_i(\mu)=\exp\left(\frac{-2}{\beta_0}\log\frac{\alpha_s(\mu)}{ \alpha_s(m_F)} \gamma_{ij} \right)
	C_j(m_F)
	\label{eq:rgsolution}
\end{flalign}
where $\beta_0=11-2/3n_f$, and $n_f=5$ is the number of running flavors.

The above procedures take into account the $\mathcal{O}(\alpha_s)$ mixing from
$O_{HG,A\tilde G}$ to $O_{tG}$.  The mixing from $O_{tG}$ to $O_{HG}$ also
exists, but is of order $\mathcal{O}(y_t^2)$, and is negligible because
$|C_{tG}|\ll |C_{HG}|$ ($C_{tG}$ is two-loop induced).  We also note here that
unlike the matching, the running in EFT is model-independent.

\subsubsection{Calculation and automation}
Once $C_i(\mu)$ is known, NLO predictions based on the EFT can be computed.  In
this work we use the automated framework based on \mg5.  Calculations at NLO
with higher-dimensional operators have been recently performed for the
top-quark sector of the SMEFT
\cite{Degrande:2014tta,Franzosi:2015osa,Zhang:2016omx,Bylund:2016phk,Maltoni:2016yxb}.
For example, in Ref.~\cite{Maltoni:2016yxb} predictions for $pp\to t\bar{t}h$
are obtained at NLO in QCD for $O_{tG}$ and the SM Higgs $O_{HG}$ operators.
The approach has the advantage that results are fully exclusive, automatically
matched to PS through {\sc MC@NLO}~\cite{Frixione:2002ik}, and can be directly used in experimental
analyses.  

A similar implementation including $O_{tG}$ and $O_{HG}$ has been created for
this study.  There are a few differences compared with previous
works:
\begin{itemize}
	\item The SM Higgs field is replaced by heavy scalar(s).\footnote{In our calculation
			we do not consider the SM Higgs contribution, $pp\to h\to t\bar t$,
			which is formally of the same order.  This contribution is largely
			suppressed as $m_{tt}\gg m_h$, and in any case it can be easily taken into
		account using the same approach.}
	\item The $O_{HG}$ (and $O_{A\tilde G}$) is defined with a prefactor of
		$g_s^2$, so that the LO signal and background have the same
		power of $\alpha_s$.  The relevant mixing is thus from $O_{HG}$
		to $O_{tG}$, not the other way around like in $pp\to t\bar{t}h$.
	\item Complex mass scheme \cite{Denner:1999gp,Denner:2005fg} is used in order to take into account the
		widths of heavy scalars.
\end{itemize}
The framework is, however, very similar and has been fully tested in previous works.

\subsection{Resolved case and reweighting}
The approach described above provides the exact NLO QCD prediction for the EFT,
which is a good approximation if the $ggH$ ($ggA$) vertex is induced by heavy
particles with masses larger than $m_{H(A)}/2$.  For lighter particles, in
particular the top quark which always couples directly to the scalar, the EFT
is not valid anymore.  In addition, it fails to capture the absorptive part of
the loop, which is crucial for understanding the lineshape.  Because of the
phase, a simple reweighting does not provide a solution. In practice, the EFT
and top-loop amplitudes become zero at different phase-space points, therefore
the amplitude ratio $|M_{\textrm{exact}}|^2/|M_{\textrm{EFT}}|^2$ used for
Born-reweighting can diverge.  As already discussed in the previous section, a
possible but ad hoc solution consists of introducing a phase in the EFT
amplitude, mimicking the absorptive part of the loop. This is achieved by
adding an imaginary part to the operator coefficient in the following simple
form $\frac{C_{HG}}{\Lambda}\to (a+b i )\times \frac{C_{HG}}{\Lambda}$
(similarly for $C_{AG}$). The values of $a$ and $b$ are constant and are
obtained at Born level by a fitting procedure that ensures that the exact and
EFT amplitudes cross zero at the same mass and with the same gradient. In more detail, 
the resolved amplitude is computed using the exact top-mass dependence in the 
form factor $A^{H,A}_{1/2}$ of Eq. \ref{eq:amplo}. The corresponding EFT result
 obtained for the infinite top mass limit, i.e.~setting $A^{H,A}_{1/2}$ to 1, is multiplied at 
 the amplitude level by $(a+b i )$.
These two forms of the amplitude are then used to compute the interference contribution to the partonic
 cross-section for $gg\to t\bar{t}$. This partonic cross section is then examined 
 as a function of the invariant mass of the top--anti-top pair. For the resolved amplitude, 
 we numerically extract the invariant mass at which the interference is zero, and also 
 compute the slope of the partonic cross-section at the same point. Requesting that the EFT result crosses
  zero at the same invariant mass, and with the same slope provides a system of 
 two equations with two unknowns which we solve to extract the value of $a$ and $b$.
This procedure ensures the ratio
$|M_{exact}|^2/|M_{EFT}|^2$ remains finite at all phase space points and allows
us to perform an event-by-event reweighting \cite{Mattelaer:2016gcx}. The
reweighting leads to the exact result at LO and a Born-improved result at NLO.
We note
here that with our setup we can also include the exact real emission amplitudes
in a fashion similar to what has been done for double Higgs production in
\cite{Frederix:2014hta,Maltoni:2014eza}, but we refrain from doing so for
simplicity. A discussion of the real emission amplitudes in the interference
for this process can be found in \cite{Hespel:2016qaf}.

\section{Benchmarks}
\label{sec:benchmarks}

To study the lineshape, we will perform the calculation for several benchmark
models, which cover both the resolved and unresolved cases. The impact of QCD
corrections in our EFT approach will be illustrated by the first two benchmark
models, where the $ggH(A)$ vertex is not resolved.  In the first scenario, we
consider a CP-even scalar that couples to a heavy vector-like fermion doublet,
which induces the $ggH$ vertex at one loop.  The matching procedure discussed
in the previous section can be explicitly carried out.  In the second model, we
consider a CP-odd scalar as a pseudo-Goldstone boson of new strong
dynamics~\cite{Chivukula:2011ue,Jia:2012kd}, whose decay into top-quarks may
shed light on an eventual dynamical fermions mass
generation~\cite{Alanne:2016rpe,Belyaev:2016ftv}.  It is important to mention
that in these two benchmarks the top-loop contributions also exist and as they
are resolved cannot be described by the EFT.  For this reason these two
benchmarks should be interpreted as ideal cases that help us to understand the
impact of NLO corrections in a scenario where the EFT approximation holds.  To
make them more realistic, we require that the top-loop contributions are always
subdominant.

For our final benchmark we consider the CP-even and CP-odd Higgses of the 2HDM.
In this case the top-loop induced $ggH$ vertex is fully resolved as only top
and bottom quarks run in the loops. For these benchmarks we construct the EFT
and improve our predictions by Born-reweighting.  

In the following we give more details for the three benchmark models.

\subsection{Benchmark A}

\begin{figure}
    \begin{center}
	\includegraphics[width=0.7\textwidth]{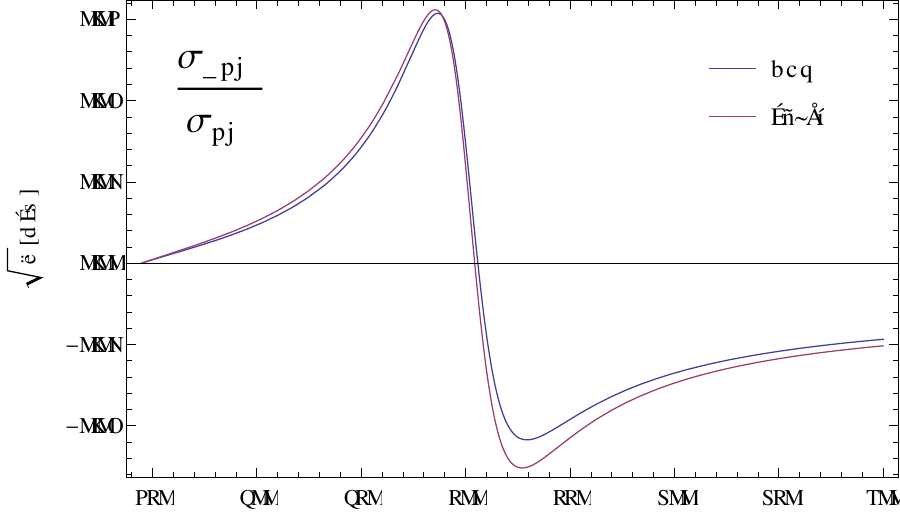}\\
    \end{center}
    \caption{Partonic cross section for $gg\to t\bar{t}$ at LO, including the
    resonance signal and the interference, normalised to the $t\bar{t}$
    background, computed from EFT and from the exact one-loop amplitude. }
       \label{fig:bench1eft}
\end{figure}

Consider a vector-like quark doublet, $F$, with Yukawa coupling to a CP-even
scalar $H$ described by Eq.~(\ref{eq:yukawa}).  We choose the following
parameters:
\begin{flalign}
	&m_H=500\ \mbox{GeV},\ 
	\Gamma_H=40\ \mbox{GeV},\
	m_F=500\ \mbox{GeV},\\
	&y_t=0.4,\ y_F=5.
\end{flalign}
The $ggH$ vertex from the $F$ fermion running in the loop is given by
Eq.~\ref{eq:2loopmatching}, multiplied by a factor of two.  For the top-quark
loop, we simply replace $m_F$ by $m_t$ and $y_F$ by $y_t/\sqrt{2}$.
This does not fully capture the top-loop contribution, in particular the phase,
however given that the dominant contribution is coming from the $F$ fermion,
the EFT is still a very good approximation.  To demonstrate this, we plot
the partonic cross sections computed from the EFT and from the exact one-loop
amplitude in Figure \ref{fig:bench1eft}.  As expected, they are quite close to each other.

Using Eqs.~(\ref{eq:2loopmatching}), (\ref{eq:2looptg}), and (\ref{eq:rgsolution}), we
find the following operator coefficients, at scale $\mu_{EFT}=m_H/2$:
\begin{flalign}
	&\frac{C_{HG}}{\Lambda}=-5.11\times10^{-5}\ \mbox{GeV}^{-1}\,,\\
	&\frac{C_{tG}}{\Lambda}=-1.40\times10^{-6}\ \mbox{GeV}^{-1}\,,
\end{flalign}
where we have taken $m_t$ to be 172.5 GeV.  The above values define our EFT. 

\subsection{Benchmark B}

\begin{figure}
    \begin{center}
	\includegraphics[width=0.7\textwidth]{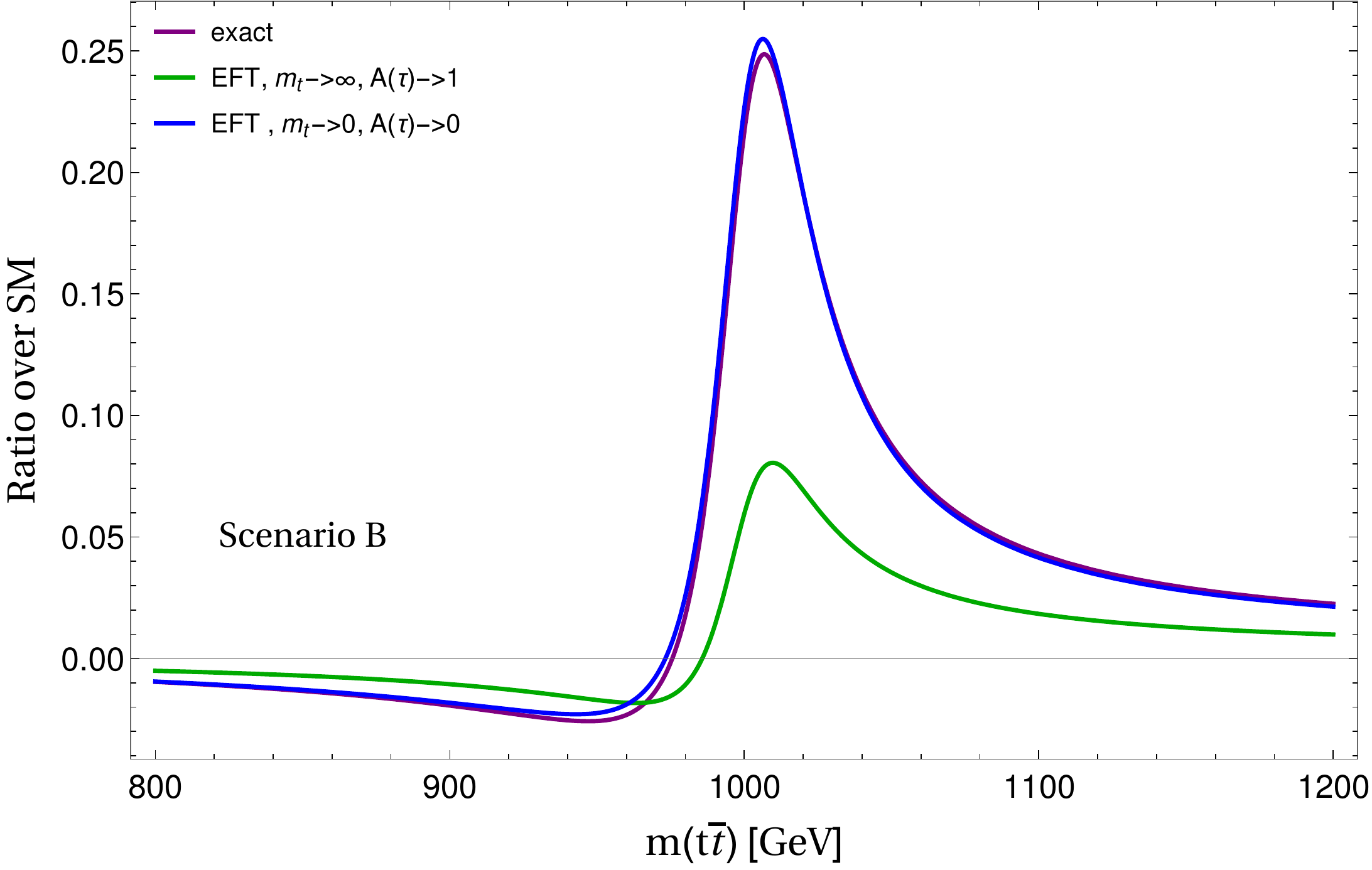}\\
    \end{center}
    \caption{Partonic cross section for $gg\to t\bar{t}$, including the
    resonance signal and the interference, normalised to the $t\bar{t}$
    background, computed from EFT and from the exact one-loop amplitude.}
       \label{fig:benchB-lineshape}
\end{figure}

As a benchmark of CP-odd state we consider the models of partial compositeness~\cite{Kaplan:1991dc} of Ref.~\cite{Belyaev:2016ftv}, more specifically the $a$ state of M3 model with $(n_\psi,n_\chi)=(-4,2)$ in the $\eta'$ decoupling limit, $\alpha=\zeta$. We choose the compositeness scale $f=800$ GeV and the mass $m_a=1$ TeV. The relevant couplings of $A\equiv a$ in \eq{eq:eff05} are
\footnote{The conversion from parameters defined in Ref.~\cite{Belyaev:2016ftv} to our convention in \eq{eq:eff05} is given by: $\frac{C_{A\tilde G}}{\Lambda}=\frac{\kappa_g}{16\pi^2 f_\pi^2}$ and $\tilde y_t = C_t \frac{m_t}{f_\pi}$}:
\begin{flalign}
	&\frac{C_{A\tilde G}}{\Lambda}=-2.15308\times 10^{-5}\ \mbox{GeV}^{-1}\,,\\
	&\tilde y_t = -0.571406 \,.
\end{flalign}
The $O_{A\tilde G}$ operator is generated through the axial-vector anomaly which couples the technipions to a pair of gauge bosons. 
We assume 
\begin{equation}
C_{tG}(1\ \mbox{TeV})=0\,.
\end{equation}
The total width is dominated by the top decay and has some correction from di-gluon decay and is given by
\begin{equation}
	\Gamma_A=37.5\ \mbox{GeV}\,.
\end{equation}
It is interesting to note that $C_{A\tilde G}$ and $\tilde y_t$ have the same
sign and this implies that the top loop has an opposite contribution to the
gluon coupling, creating a dip-peak structure instead of the usual peak and
dip.  This opposite contribution also generates some strong cancellations if
one approximates the full top loop with a heavy fermion, as we have done for
benchmark A, thus for this benchmark we use only the effective coefficient
generated by the high scale physics and neglect the top loop in the gluon
coupling. This is illustrated in Figure~\ref{fig:benchB-lineshape}, where the
partonic cross section normalised to the QCD background computed at LO is
displayed.  The difference between the curves stems from the way to treat the
resolved top loop in the amplitude in \eq{eq:amplo}.  The exact curve has the
full form factor structure. The heavy fermion limit $m_t\to \infty$ and
$A^A_{1/2}(\tau)\to 1$ is a bad approximation because of the large cancellation
between $c_g$ and $c_t$.  A better approximation, given by the blue curve, is
to use the limit where $m_t\to 0$ and the top loop vanishes, which is justified
by the heavy scales we are probing. 

\subsection{Benchmarks C1 and C2}
\label{sec:model} 
For our final BSM scenario we employ the 2HDM~\cite{Branco:2011iw}, which introduces  
a second $SU(2)_L$ doublet $\Phi_2$ and gives rise to five physical
Higgs bosons: one light (heavy) neutral, CP-even state $h$ ($H$); 
one neutral, CP-odd state $A$; and two charged Higgs bosons $H^{\pm}$.
In this work we take $h$ to be the 125 GeV Higgs.
The input parameters determining all properties of a 2HDM scenario are:
\begin{alignat}{9}
 \tan\beta\;, \sin\alpha\;, m_{h}\;, m_{H}\;, m_{A}\;, m_{H^{\pm}}\;, m^2_{12}
 \label{eq:freeinputs},
\end{alignat}
with the convention
$0 \leq \beta-\alpha < \pi$ (with $0 < \beta < \pi/2$). 

The ratio of the Yukawa couplings over the SM values are shown 
in Table~\ref{tab:yukawas} for both type--I and type--II 2HDMs. 
  \begin{table}[t!]
 \centering
\begin{tabular}{l|l}
 \multicolumn{2}{c}{Type--I and II}  \\ \hline
 $1+\Delta^{h}_t$ & $\cfrac{\cos\alpha}{\sin\beta} $  \\
 $1+\Delta^{H}_t$ & $\cfrac{\sin\alpha}{\sin\beta}  $  \\
 $1+\Delta^{A}_t$ & $\cot \beta$  \\
\end{tabular}
\caption{Top quark Yukawa couplings to the light (heavy)
CP-even and CP-odd Higgs bosons. These are identical for type--I and type--II
2HDM. \label{tab:yukawas}}
\end{table}
Electroweak precision tests, the LHC Higgs results and searches for heavy
scalar particles, along with unitarity, perturbativity and vacuum stability
constrain the parameter space of the model. In the selection of 2HDM
benchmarks, these constraints are taken into account included through the
public tools { \sc 2HDMC}~\cite{Eriksson:2009ws}, { \sc
HiggsBounds}~\cite{Bechtle:2008jh,Bechtle:2011sb} and {\sc{HiggsSignals}}
\cite{Bechtle:2013xfa,Stal:2013hwa}. The two benchmarks we will employ are
shown in Table \ref{tab:benchmarks} and are briefly discussed below.

\begin{table}[tb!]
\begin{center}
 \begin{tabular}{|lc|rc|rrrr|} \hline 
 &Type & $\tan\beta$ & sin($\beta-\alpha$) & $m_{H} $ &  $m_{A} $  & $m_{H^{\pm}} $   & $m^2_{12} $  \\ \hline
C1 & I & 2.0 & 1.0 & 300 & 450 & 450 & 20000 \\
C2 & II & 0.9 & 1.0 & 450 & 600 & 620 & 10000 \\
\hline
 \end{tabular}
 \end{center}
 \caption{Parameter choices for the 2HDM benchmarks used in our study. All
 masses are given in GeV. The light Higgs mass is fixed to $m_{h} = 125$ GeV.  \label{tab:benchmarks}}
\end{table}
 
\subsubsection{C1: CP-odd resonance}

For this benchmark only the pseudoscalar Higgs lies above the top--anti-top threshold. The corresponding Yukawa couplings (as rescalings of the SM Higgs top Yukawa) are: $g^H_t=-0.5$ and $g^A_t=0.5$. The heavy scalar width is negligible while the pseudoscalar width is $\Gamma_A=7.35$ GeV and the top quark decay branching fraction is approximately 65\%.   

\begin{figure}[h]
    \begin{center}
	\includegraphics[width=0.9\textwidth,trim=0cm 15cm 5cm 4cm]{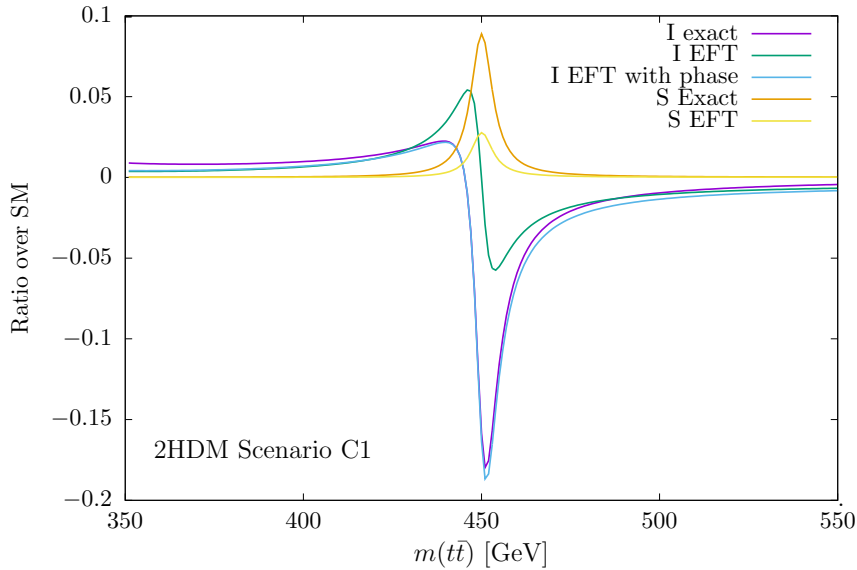}
    \end{center}
    \caption{Signal (S) and interference (I) over background ratios, 
    computed from EFT (with and without the additional phase) and from the exact one-loop amplitude.}
       \label{fig:B1phase}
\end{figure}

Following Eqs.~(\ref{eq:2loopmatching}), (\ref{eq:2looptg}), and (\ref{eq:rgsolution}), we
find the following operator coefficients, at scale $\mu_{EFT}=m_A/2$:
\begin{flalign}
	&\frac{C_{HG}}{\Lambda}=4.73\times10^{-6}\ \mbox{GeV}^{-1}\,,\\
	&\frac{C_{AG}}{\Lambda}=-6.49\times10^{-6}\ \mbox{GeV}^{-1}\,,\\
	&\frac{C_{tG}}{\Lambda}=9.56\times10^{-9}\ \mbox{GeV}^{-1}\,.	
\end{flalign}
Where we have taken $m_t$ to be 172.5 GeV.  The above values define our EFT.  For this scenario, the EFT is not expected to accurately describe the lineshape as the top-loop is resolved. The ratio over the SM background is shown at LO in Figure \ref{fig:B1phase} for the interference (I) and signal (S). In particular the interference lineshape differs between the one-loop and EFT calculation, due to the phase difference between the amplitudes. A solution is to introduce a complex phase, in the ad-hoc form of  $\frac{C_{AG}}{\Lambda}\to (a+b i )\times \frac{C_{AG}}{\Lambda}$, where $a$ and $b$ are extracted by matching the exact and EFT amplitudes at leading order in particular at the point where they cross zero. For this benchmark introducing this phase can be used to improve the description of the lineshape as shown in Figure \ref{fig:B1phase}. This is particularly useful also for the NLO computation which is improved by Born reweighting. 

\subsubsection{C2: CP-even and CP-odd resonances}

The corresponding couplings and widths for this benchmark are: $g^H_t=-1.11$, $g^A_t=1.11$, $\Gamma_H=10.7$ GeV and $\Gamma_A=38.7$ GeV. Both resonances decaying almost exclusively to top quarks with the relevant branching fractions being $Br(H\to t\bar{t})=0.995$ and $Br(A\to t\bar{t})=0.921$.

\begin{figure}[h]
    \begin{center}
	\includegraphics[width=0.8\textwidth,trim=0cm 15cm 5cm 4cm]{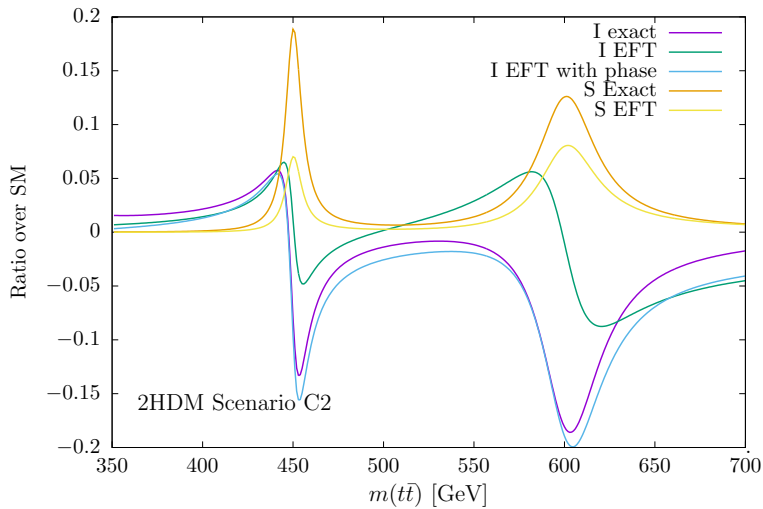}
    \end{center}
    \caption{Signal and interference over background ratios, 
    computed from EFT (with and without the additional phase) and from the exact one-loop amplitude.}
       \label{fig:B2phase}
\end{figure}

Following Eqs.~(\ref{eq:2loopmatching}), (\ref{eq:2looptg}), and (\ref{eq:rgsolution}), we
find the following operator coefficients, at scale $\mu_{EFT}=m_A/2$:
\begin{flalign}
	&\frac{C_{HG}}{\Lambda}=1.05\times10^{-5}\ \mbox{GeV}^{-1}\,,\\
	&\frac{C_{AG}}{\Lambda}=-1.44\times10^{-5}\ \mbox{GeV}^{-1}\,,\\
	&\frac{C_{tG}}{\Lambda}=4.11\times10^{-8}\ \mbox{GeV}^{-1}\,.	
\end{flalign}
The above values define our EFT. Similarly to benchmark C1 the EFT does not provide a reliable prediction of the lineshape and a phase (one for the scalar and one for the pseudoscalar resonance) can be used in order to match the lineshape of the interference. The results are shown in  Figure \ref{fig:B2phase}.

\section{Results}
\label{sec:results}
In this section we present results for the four benchmarks described in section
\ref{sec:benchmarks}.  We fix our EFT scale at one half of the scalar resonance mass,
while for renormalisation and factorisation scales we choose a dynamical scale
equal to one half of the sum of the transverse masses in the final state.
We vary the scales independently by a factor of two up and down to estimate the
scale uncertainties.  The Yukawa coupling between the $H$ and the top quark
is renormalised by the $\overline{MS}$ scheme but fixed at the scale $m_H/2$.  We
use the LO and NLO NNPDF2.3 parton distribution functions \cite{Ball:2013hta}
for the corresponding computation. We obtain results both at fixed-order and
also matched to the parton shower with MC@NLO. For the parton shower we always use
\Pythia8 \cite{Sjostrand:2007gs}.
The fully inclusive phase-space is considered and the tops are kept stable, even though their decays can be considered in a fully automatic way via the \textsc{MadSpin}~\cite{Artoisenet:2012st} package allowing further analysis of exclusive and realistic observables. 

\subsection{Benchmark A}
\begin{table}[htb]
	\centering
	\begin{tabular}{cccc}
	\hline\hline
	& LO & NLO & $K$-factor 
	\\\hline
	SM & $473.9^{+29\%}_{-22\%}$ & $685.0^{+10\%}_{-12\%}$
	& $1.45^{+13\%}_{-15\%}$
	\\
	Interference & $-1.51^{+25\%}_{-37\%}$ & $-1.50^{+9\%}_{-6\%}$
	& $1.00^{+22\%}_{-25\%}$
	\\
	Signal & $2.36^{+30\%}_{-23\%}$ & $4.23^{+18\%}_{-16\%}$
	& $1.79^{+9\%}_{-10\%}$
	\\\hline\hline
	\end{tabular}
	\caption{\label{tab:A}Cross sections for benchmark A in pb.  Uncertainties
	are from renormalisation and factorisation scale variation.}
\end{table}

\begin{figure}
    \begin{center}
	\includegraphics[width=0.7\textwidth]{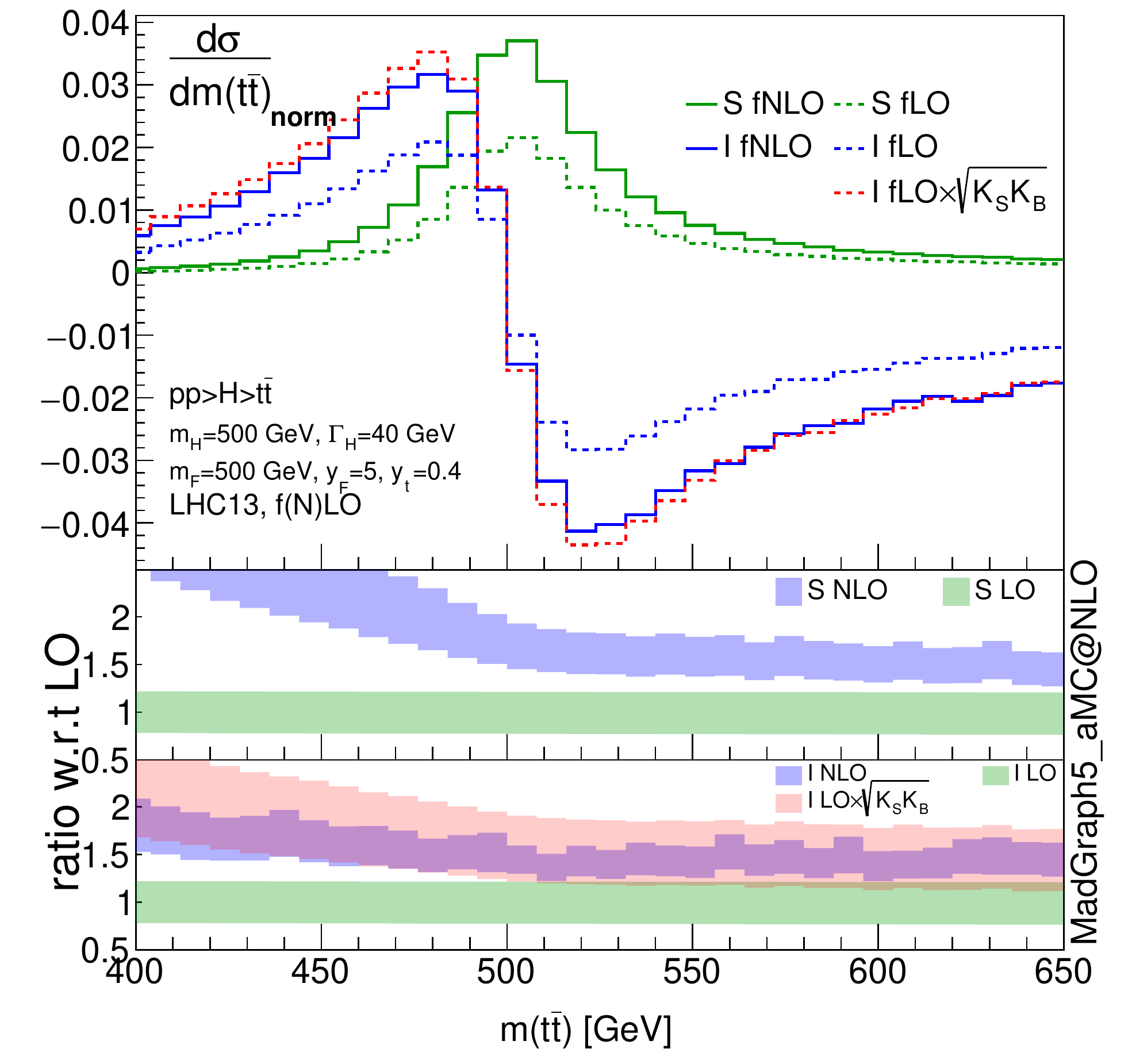}
    \end{center}
    \caption{Signal and interference lineshapes at fixed order (N)LO for LHC 13 TeV.
    Results are normalised to the SM fixed order at NLO (fNLO) lineshape.
    Lower panels show the K-factors and the scale uncertainties of signal
    and interference respectively.}
       \label{fig:A1}
\end{figure}

\begin{figure}
    \begin{center}
	\includegraphics[width=0.7\textwidth]{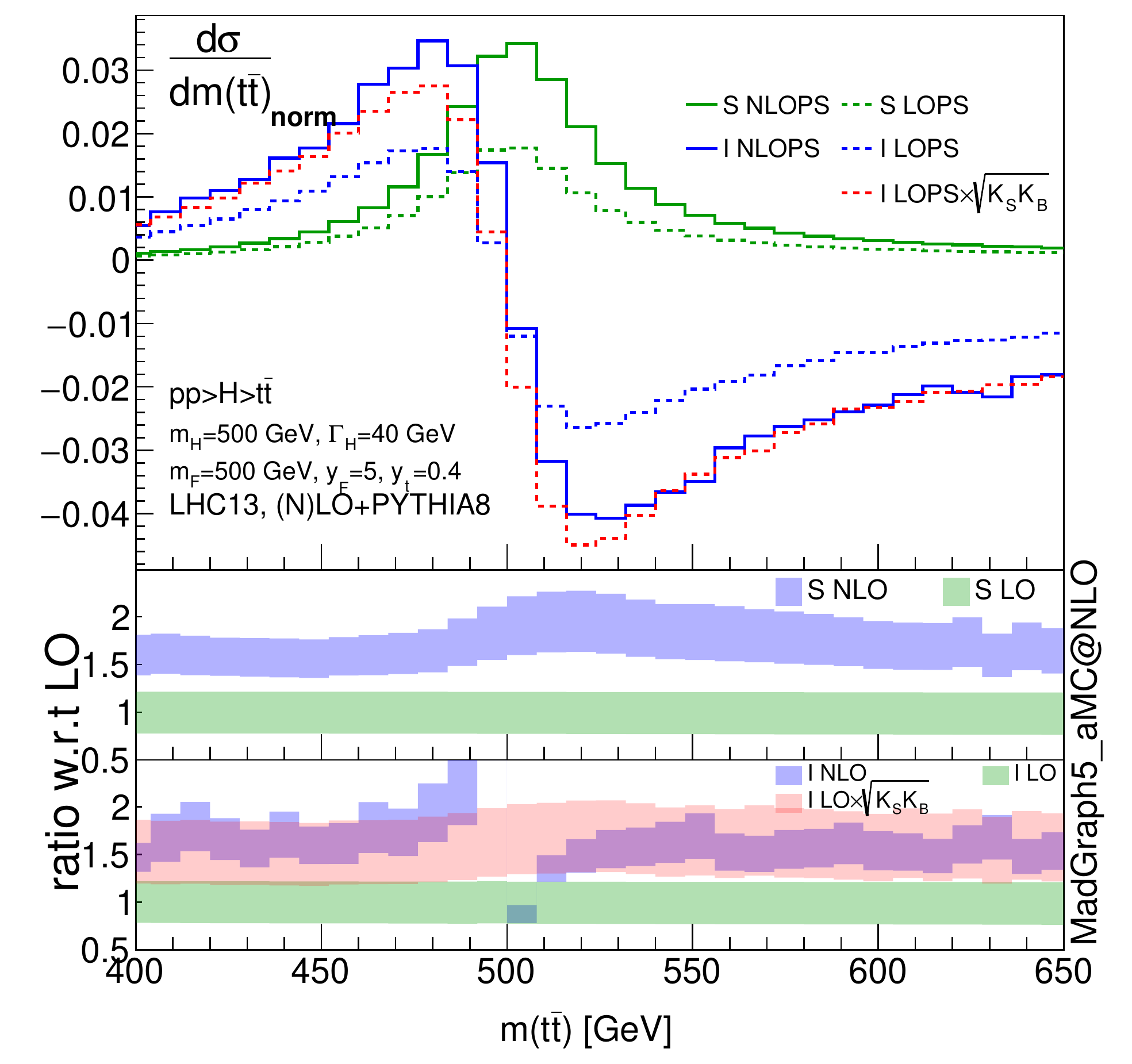}
    \end{center}
    \caption{Signal and interference lineshapes at (N)LO matched with PS for LHC 13 TeV.
    Results are normalised to the SM fNLO lineshape.
    Lower panels show the K-factors and the scale uncertainties of signal
    and interference respectively.}
       \label{fig:A2}
\end{figure}

The cross sections for the SM background, the interference, and the signal
components at the LHC 13 TeV are given in Table~\ref{tab:A}.  Large QCD
corrections can be observed for the signal, but not for the interference.  This
however does not mean that the radiative corrections are not important for the
interference, as the interference displays a cancellation below and above the
resonance.  In fact, looking at the scale uncertainties, we can see that our
calculation improves the precision by a factor of $\sim4$.

To further investigate our results, in Figure~\ref{fig:A1} we show the fixed
order $m_{t\bar{t}}$ distribution.  Signal (S) and interference (I) lineshapes are displayed
separately.  The signal peaks at the resonant mass 500 GeV, and the radiative
correction is quite large as expected. In particular, we observe an increasing
$K$-factor as the $t\bar t$ invariant mass moves away from the resonance to lower values.  This is
mainly because the heavy Higgs decays to $t\bar t+g$, where the gluon reduces
the total energy of the $t\bar t$ pair, so that events near the peak at LO may be
shifted to the left at NLO.  A similar but less significant effect can be
observed also on the interference lineshape, which shows a typical peak-dip
structure.  In the lower panels, we find that radiative corrections to the rate
are large, while the improvements on the scale uncertainties are mild.
Finally, as a comparison, we follow the approach in Ref.~\cite{Hespel:2016qaf}
and rescale the LO lineshape by a $K$-factor inferred by using
$\sqrt{K_SK_B}$, where $K_S$ and $K_B$ are the (bin by bin) $K$-factors for
the SM and the signal. As shown in the plot, the resulting lineshape
is a reasonable approximation in the absence of a full NLO computation for
the interference which we provide in this work, but it tends to overestimate
the size of the interference at and below the resonance.  In addition, by construction 
this approach does not improve the LO scale uncertainty, as can be observed
in the last panel.

In Figure~\ref{fig:A2}, the same results are shown but matched to the PS. Compared with
Figure~\ref{fig:A1}, the size of the radiative corrections is now tamed below the
resonance, mainly because the emission from decay products is partly taken into
account by the PS.  The $K$-factor for the signal is in general flat, with a small
enhancement near the peak.  In contrast, the $K$-factor for the interference
shows a peak-dip structure, mainly because the zero point has been slightly shifted.
This effect is not reproduced by a $K$-factor rescaling used in
Ref.~\cite{Hespel:2016qaf}.  As one can see in the last panel, the inferred
$K$-factor $\sqrt{K_SK_B}$ is flat and does not reproduce the
peak-dip structure, and as a result this approximation will predict an
interference lineshape that is shifted to the left compared with our full
calculation.  Finally, compared with the fixed-order results, the showered
lineshape has a slightly lower peak due to smearing effects.

\begin{figure}
    \begin{center}
	\includegraphics[width=0.7\textwidth]{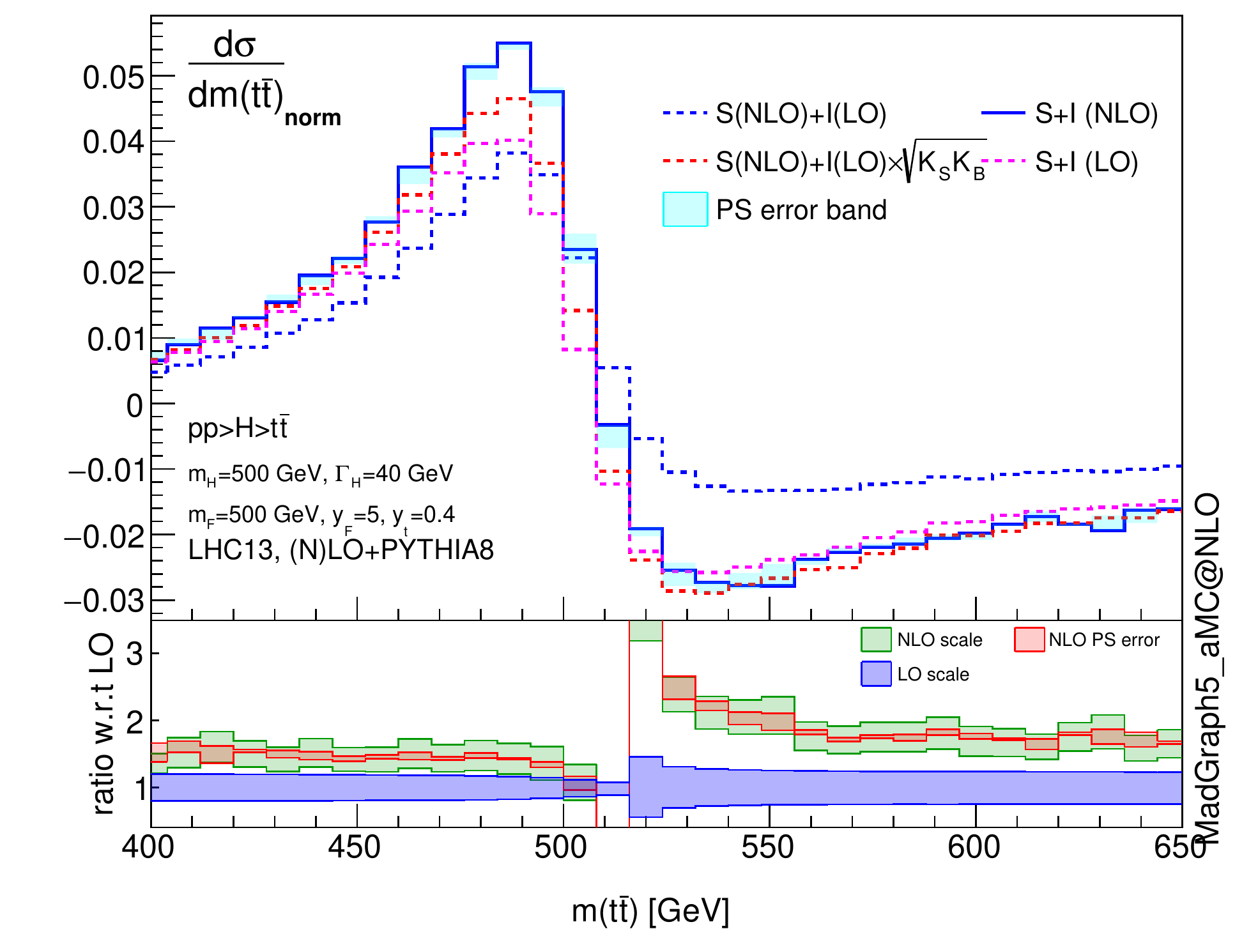}
    \end{center}
    \caption{Total BSM effects (signal+interference) with the interference
	    at LO/NLO, at the LHC 13 TeV matched to PS.
    Results are normalised to the SM fNLO lineshape,
    except for the LO curve which is normalised to SM fLO.
    The lower panel displays the K-factors and the scale uncertainties of signal
    and interference respectively.}
       \label{fig:A3}
\end{figure}

In Figure~\ref{fig:A3}, we show the total BSM effect, including signal and
interference.  In particular, we compare the sum of NLO signal and LO
interference with the full NLO prediction, and their ratio is given in the
lower panel.  The comparison shows the impact of this work.  The rate below the
resonance is increased by about 50\%, while the increase above the resonance is
even larger.  A shift of the zero point is also observed.  Again, we compare
with the $K$-factor rescaling approximation used in Ref.~\cite{Hespel:2016qaf}.
Similar to Figure~\ref{fig:A2}, we find that this approximation predicts a
lower lineshape, i.e.~it tends to underestimate the peak, but overestimate the
dip. It also underestimates the resonant mass of the scalar.
In addition, in the same plot we show the sum of LO signal and LO interference,
normalised to SM LO background.  The full NLO
results normalised to NLO background gives a much higher peak, which is
expected because the QCD correction to the resonance is larger than that to the
background.

In the same graph we also plot the error due to the PS treatment of the interference.
The origin of the error is explained as follows.  In general, the PS algorithm
as implemented in {\sc Pythia8} distinguishes between events with and without an
explicit resonant state. If the resonance is present, showering is treated in a
factorised way.  External coloured legs of the production part are generated,
in such a way so that the invariant mass of the resonance is conserved.  If an
intermediate resonance is not present, then the PS only sees the full process,
and external coloured legs will be generated without conserving the invariant
mass of the resonance.  Another difference, particular to this process, is that
events with and without a colour neutral resonance correspond to different
classical colour flows.  Both facts affect the results of showering.  In
practice, we can treat the SM background as non-resonant while the signal part
as resonant. However, whether the interference events should be considered with
a resonant intermediate state is not well-defined, and this is the origin of
what we call PS uncertainty of the interference.  The problem is not so severe
for processes with small interference contributions.  However, the process
we consider here is a special one, where it is the interference contribution that
really determines the shape of the BSM contribution.  Therefore it is important
to assess the possible consequences of this uncertainty. 

By default, samples generated by {\sc MG5\_aMC} contain both resonant and
non-resonant events, depending on the relative size of signal and background.
Such a scheme is not a physical one, as it implies that the shape of the
interference will be determined by the size of the effective operator coefficient.  To
estimate the uncertainty due to this fact, we generate two additional sets of
events, one with only resonant events, the other with only non-resonant. We then
pass these two sets of events through the PS, and the resulting $m_{t\bar t}$
distributions form an error band, which is displayed in Figure~\ref{fig:A3} by
a light blue band.  We can see that this effect is not important away from the
resonance, however it does shift the lineshape near the resonant area, where the 
interference is expected to be large.  The normalised uncertainty is given in
the lower panel by the red band.  Again we see that in general this
uncertainty is small, except for the near-resonance region, where this effect
can become comparable to the scale uncertainties.

In Refs~\cite{Frederix:2016rdc} and \cite{Jezo:2015aia,Jezo:2016ujg} resonance aware matching has been proposed within the MC@NLO and {\sc Powheg} frameworks respectively, with focus on narrow coloured resonances.
In our configuration the resonances are broader and the interference plays a more important role. In the case of the large interference the ambiguities related to the treatment of the resonance are more severe, and our estimate of the parton shower uncertainty could well be an underestimate. 

In general the PS uncertainty arises from the intrinsic limitations of the current PS algorithms
that are publicly implemented, and therefore a solution to this problem is beyond
the scope of this work.  We estimate its potential impact to benchmark A,
but we do not expect a significant difference for the other benchmarks.
For this reason we will not consider this error for the rest of the paper.


\subsection{Benchmark B}

\begin{table}[htb]
	\centering
	\begin{tabular}{cccc}
	\hline\hline
	& LO & NLO & $K$-factor
	\\\hline
	SM & $44.7^{+35\%}_{-25\%}$ & $61.4^{+13\%}_{-14\%}$ & $1.37^{  +  14\%}_{  -  17\%}$
	\\
	Interference & $-0.24^{+23\%}_{-32\%}$ & $-0.39^{+16\%}_{-17\%}$ & $1.66^{ +  10\%}_{  -  12\%}$
	\\
	Signal & $0.61^{+36\%}_{-25\%}$ & $0.90^{+16\%}_{-16\%}$ & $1.48^{  +  12\%}_{  -  14\%}$
	\\\hline\hline
	\end{tabular}
	\caption{\label{tab:B}Cross sections for benchmark B in pb with $m(t\bar{t})>750$ GeV.  Uncertainties
	are from renormalisation and factorisation scale variation.}
\end{table}

\begin{figure}
    \begin{center}
	\includegraphics[width=0.7\textwidth]{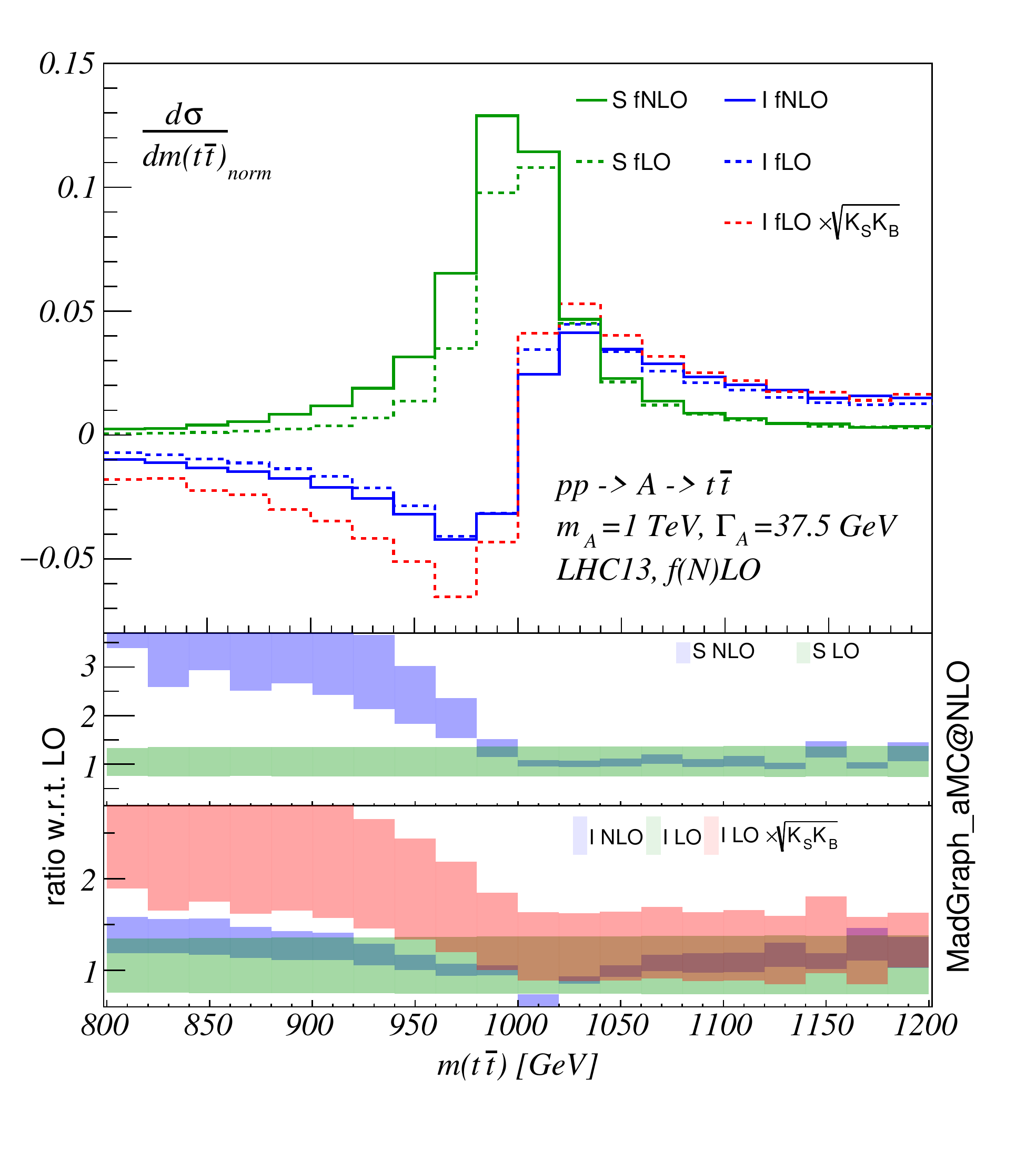}
    \end{center}
    \caption{Signal and interference lineshapes at fixed order (N)LO for LHC 13 TeV.
    Results are normalised to the SM fNLO lineshape.
    Lower panels show the K-factors and the scale uncertainties of signal
    and interference respectively.}
       \label{fig:B1}
\end{figure}

\begin{figure}
    \begin{center}
	\includegraphics[width=0.7\textwidth]{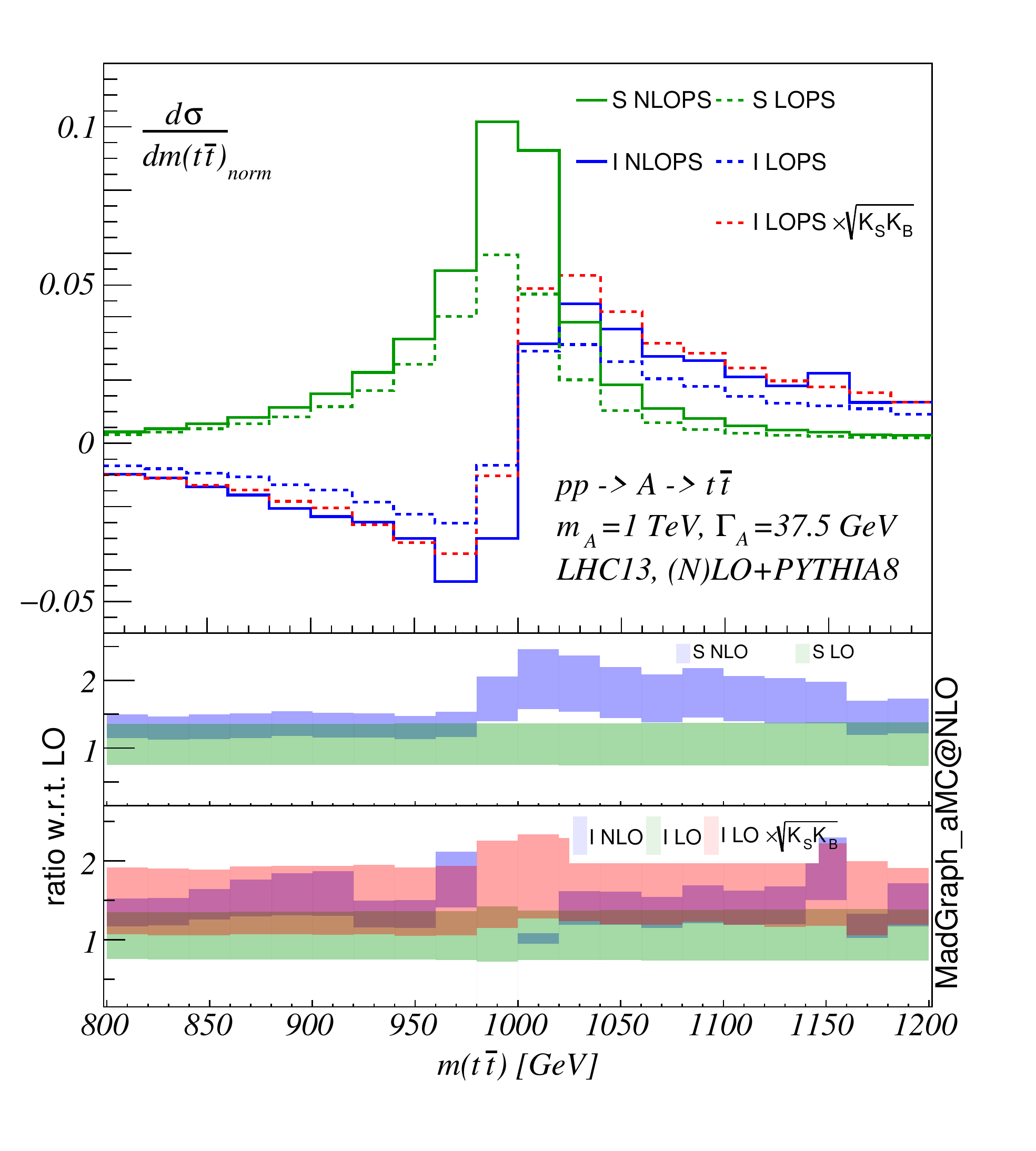}
    \end{center}
    \caption{Signal and interference lineshapes at (N)LO matched with PS for LHC 13 TeV.
    Results are normalised to the SM fNLO lineshape.
    Lower panels show the K-factors and the scale uncertainties of signal
    and interference respectively.}
       \label{fig:B2}
\end{figure}

The cross section for Benchmark B is shown in \tab{tab:B} with a generation cut
applied to the invariant mass of the top pair system, $m(t\bar{t})>750$ GeV.
This is to ensure that we have enough statistics in the high $m(t\bar{t})$
region near the resonance.

In \fig{fig:B1} we show the invariant mass distribution of the top pair system
$m_{t\bar{t}}$ at fixed order in perturbation theory. In the upper panel the
NLO and LO predictions are shown normalised bin-by-bin to the background QCD
prediction. Pure signal (S) and interference (I) are shown separately. Results
obtained using the approximate average $K$-factor, $\sqrt{K_SK_B}$, are also
shown. The lower panels show the $K$-factors and the scale uncertainty as an
envelope.   For the signal, the same behaviour observed for the scalar
benchmark can be noticed: a large effect at the lower tail of the distribution
can be observed as the gluon emission reduces the energy of the $t\bar{t}$
system. In the high energy tail a smaller scale uncertainty is obtained with
the NLO calculation.  For the interference, the QCD corrections increase the
rates mainly below the resonance for the fixed-order predictions. It is
interesting to restate that in this scenario there is a dip-peak structure due
to the opposite sign contribution of the high energy strong sector. The
approximate $K$-factor, $\sqrt{K_SK_B}$, overestimates the size of the
corrections, in particular in the region below the resonance. 

In Figure \ref{fig:B2} we show the equivalent distribution with matching to the
PS. The conclusions are similar to those described in the previous benchmark.
The signal $K$-factors peaks at the resonance mass, while the interference
$K$-factor is mostly flat for the NLO+PS predictions.   The overall $K$-factors
are however lower than in benchmark A, due to the higher mass considered. 

\subsection{Benchmarks C1 and C2}

\begin{table}[htb]
	\centering
	\begin{tabular}{cccc}
	\hline\hline
	& LO & NLO & $K$-factor 
	\\\hline
	SM & $473.9^{+29\%}_{-22\%}$ & $685.0^{+10\%}_{-12\%}$ & $1.45^{+13\%}_{-15\%}$
	\\
	Interference & $-1.64^{+34\%}_{-25\%}$ & $-2.30^{+14\%}_{-14\%}$ & $1.40^{+14\%}_{-15\%}$
	\\
	Signal & $1.15^{+32\%}_{-24\%}$ & $1.98^{+18\%}_{-16\%}$ & $1.72^{+10\%}_{-11\%}$
	\\\hline\hline
	\end{tabular}
	\caption{\label{tab:C1}Cross sections for benchmark C1 in pb.  Uncertainties
	are from renormalisation and factorisation scale variation.}
\end{table}

\begin{table}[htb]
	\centering
	\begin{tabular}{cccc}
	\hline\hline
	& LO & NLO & $K$-factor 
	\\\hline
	SM & $473.9^{+29\%}_{-22\%}$ & $685.0^{+10\%}_{-12\%}$ & $1.45^{+13\%}_{-15\%}$	\\
	Interference & $-3.35^{+37\%}_{-26\%}$ & $-3.84^{+8\%}_{-11\%}$ & $1.14^{+19\%}_{-21\%}$	\\
	Signal & $5.52^{+33\%}_{-24\%}$ & $8.91^{+12\%}_{-16\%}$ & $1.61^{+12\%}_{-13\%}$
	\\\hline\hline
	\end{tabular}
	\caption{\label{tab:C2}Cross sections for benchmark C2 in pb.  Uncertainties
	are from renormalisation and factorisation scale variation.}
\end{table}

The cross sections for the signal, background and their interference for
scenarios C1 and C2 are shown in Table \ref{tab:C1} and \ref{tab:C2}
respectively.  The QCD corrections change significantly the central value of
the prediction but also reduce the scale uncertainties. 
\begin{figure}[h]
    \begin{center}
	\includegraphics[width=0.6\textwidth,trim=2cm 3cm 2cm 2cm]{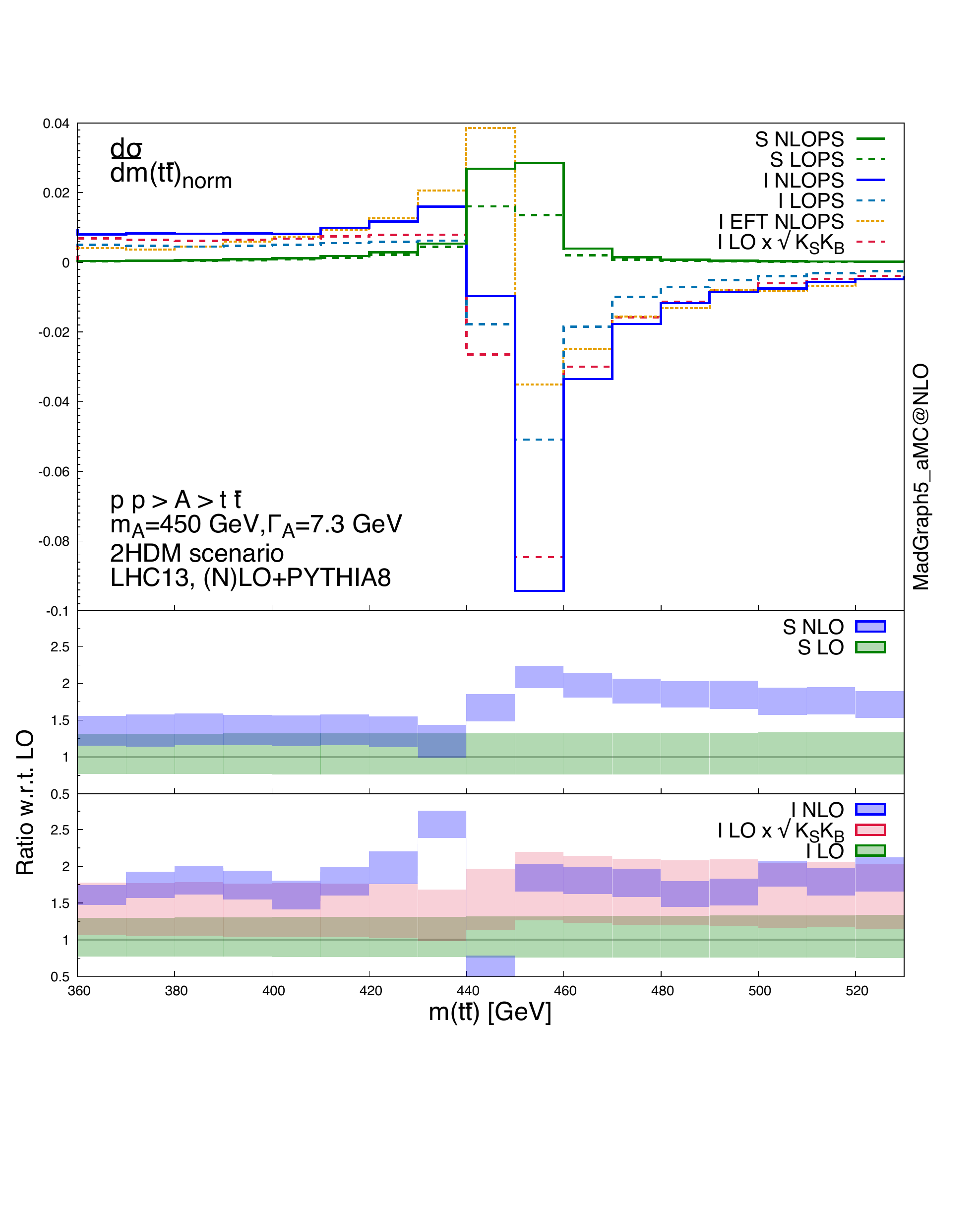}
    \end{center}
    \caption{Signal and interference lineshapes at (N)LO matched with PS for LHC 13 TeV.
    Results are normalised to the SM fNLO lineshape.
    Lower panels show the K-factors and the scale uncertainties of signal
    and interference respectively. }
       \label{fig:C1}
\end{figure}

\begin{figure}[h]
    \begin{center}
	\includegraphics[width=0.6\textwidth,trim=2cm 3cm 2cm 2cm]{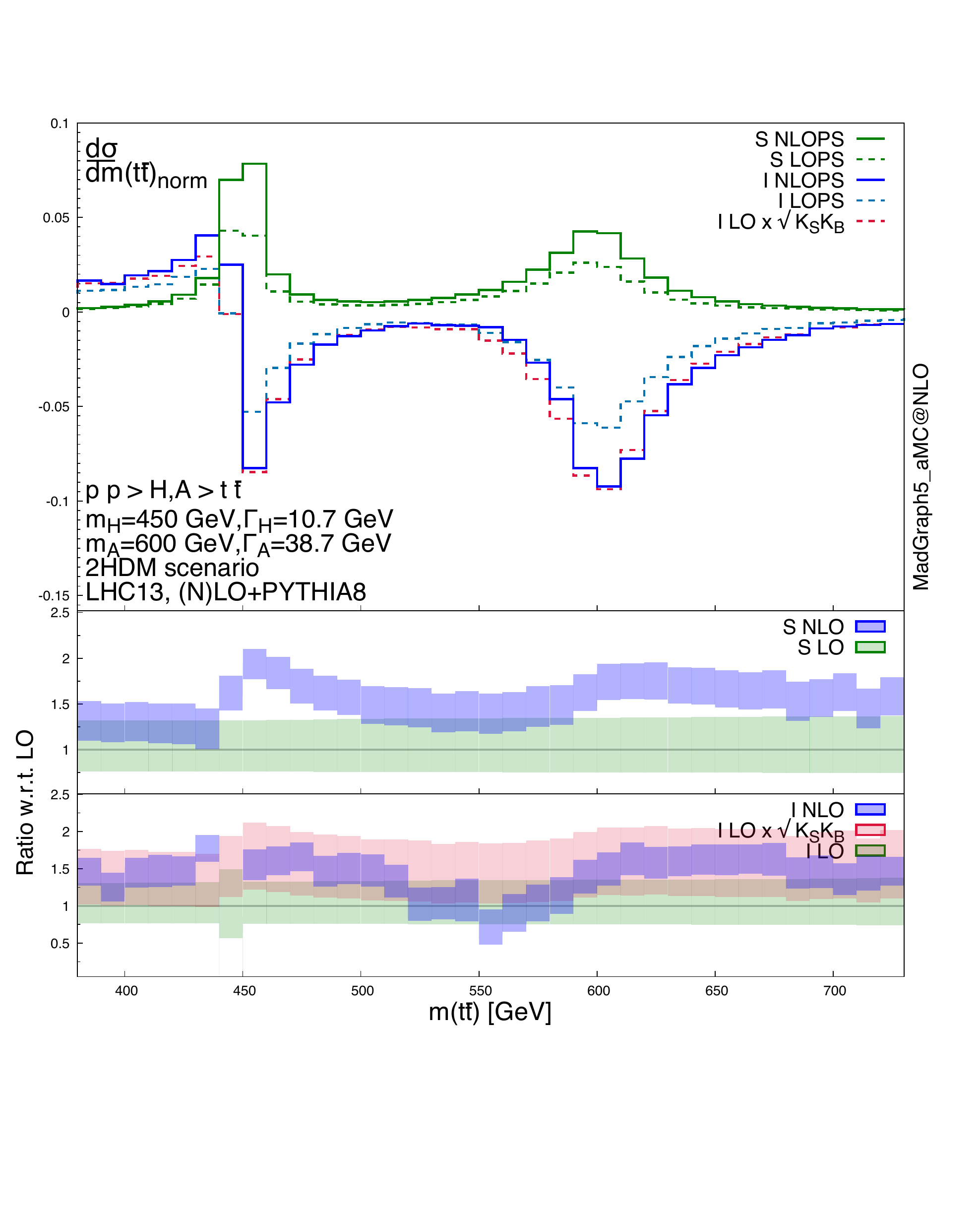}
    \end{center}
    \caption{Signal and interference lineshapes at (N)LO matched with PS for LHC 13 TeV.
    Results are normalised to the SM fNLO lineshape.
    Lower panels show the K-factors and the scale uncertainties of signal
    and interference respectively.}
       \label{fig:C2}
\end{figure}

Differential results for the 2HDM benchmarks C1 and C2 are shown in Figures
\ref{fig:C1} and \ref{fig:C2} respectively. The invariant mass distribution of
the top quark pair is computed at (N)LO+PS, with the signal (S) and interference (I)
contributions and their corresponding $K$-factors displayed separately.  The LO
results are exact, i.e.~they include the full top mass dependence, while for
the NLO results we use the phase-improved EFT to compute the NLO corrections
and then apply a Born-reweighting at an event-by-event level, as discussed in
the previous section. As the reweighting is based on event generation we
present only (N)LO+PS for these benchmarks. Results obtained using the
approximate average $K$-factor, $\sqrt{K_SK_B}$, are also shown. For both
benchmarks the signal $K$-factors peak around the resonant masses. The
interference $K-$factor shows a peak-dip structure as the NLO corrections shift
the zero crossing point from its LO position. The approximate  $K$-factor,
$\sqrt{K_SK_B}$ is much flatter, with small peaks driven by the peaks in the
signal $K$-factor at the resonance masses. Moreover, the uncertainties of that
prediction are much larger as they are by construction LO. 

For benchmark C1 we also show in yellow the prediction obtained for the
interference in the EFT limit, i.e.~without any reweighting. This result should
be close to the approximation in \cite{Bernreuther:2015fts}. As expected
from the partonic results shown in Figure \ref{fig:B1phase} the EFT prediction
does not a provide a good description of the interference lineshape. 

We stress that for scenarios where two resonances are present the experimental
resolution will be crucial to establish whether one or two peaks are visible in
the invariant mass distribution.  For our benchmark the mass difference of 150
GeV should be sufficiently large.  We note that here the interference
distribution suffers from low statistics in the region between the two
resonance masses, where the cross-section is very small.

\section{Conclusions}
\label{sec:conclusion}
We have computed for the first time the NLO QCD corrections for the
interference between signal and background in resonant scalar or pseudoscalar
top pair production using an EFT approach. The interference between signal and
background is crucial in the determination of the resonant lineshape in a
series of motivated BSM scenarios, where new heavy scalar and/or pseudoscalar
particles are coupled to the top quarks. In the case where the gluon fusion
scalar production is dominated by heavy particles running in the loop, or
strong dynamics introducing a point-like interaction between the scalar and the
gluons, the EFT provides a very good approximation of the process, which we
have verified by analytical calculation. The computation at NLO in the EFT
limit reduces the problem to one-loop level and can be performed without further
approximations. 

Using this approach, we have computed the NLO corrections within the \mg5
framework, to obtain an accurate description of the $t\bar t$ lineshape.
As a nontrivial feature of this calculation, we found that the effective
gluon-scalar operator mixes into the chromo-magnetic dipole operator, which
implies that both operators, together with their mixing effects, need to be
incorporated in the computation to guarantee a finite and physical result. To
demonstrate how the coefficients of both operators can be extracted from the
underlying theory, we have performed two-loop matching to a UV complete model,
and evolved the operator coefficients to the scale where the calculation will
be performed. Whilst the matching is model dependent, the running and mixing
between the operators is model independent.  Furthermore, to handle the cases
where the EFT is less applicable, i.e.~when there are light particles running
in the gluon fusion loop, we have demonstrated that the EFT results can still
be used to improve the predictions.  This can be done by introducing a complex
phase in the coefficient of our effective operators to match the absorptive
part of the one-loop amplitude, and employing a Born-reweighting using the
exact LO amplitudes to obtain results beyond LO. 

Our setup is fully automated and allows us to compute results both at the fixed
order and matched to the parton shower. We have presented results for a
representative set of benchmarks, covering both scalar and pseudoscalar
resonances in the unresolved and resolved cases. In particular, we have studied
a scenario with a scalar resonance coupling to a heavy vector-like quark
doublet, a scenario of a pseudoscalar state in a model of partial
compositeness, and two benchmarks in the 2HDM. The first two scenarios can be
well described by the EFT, whilst for the 2HDM scenarios, a complex phase and
reweighting have been used to improve the EFT predictions. In all cases we have
found that QCD corrections are important and significantly reduce the scale
uncertainties of the predictions. At the differential level the corrections
lead to nontrivial $K$-factors and therefore are crucial for a precise
determination of the lineshape. We examined both fixed-order and NLO+PS
distributions, and in particular assessed the impact of the uncertainty due to
the treatment of the interference contribution in the matching to the PS
simulation. We found that this uncertainty is relevant in the region close to
the resonance where it can become comparable to the scale uncertainties.
Finally, we compared our predictions with previous approximate NLO results,
which are based on the geometric mean of the respective signal and background
$K$-factors. We found significant improvements in the resonant region. 

In summary, we have computed the NLO QCD corrections to resonant scalar
production and decay into top quarks in the EFT limit taking into account the
signal and background interference. Our computation improves the accuracy and
precision of the theory predictions, allowing for more reliable analyses of
resonant top pair production. Matching to the PS simulations is provided in an
automated way, and thus predictions can be used in realistic simulations in the
context of resonant searches at the LHC. Combined with advances in the
experimental techniques, we expect that our work will improve the sensitivity of
experimental searches for BSM scalars, by reducing the systematical uncertainty
from the theory side, and by optimising the experimental strategies according
to an accurate lineshape description. On the theory side, it will also allow us
to extract reliable constraints on the parameters of various BSM models.
Finally, the calculation itself is an interesting one, as the two-loop matching
to both operators allows for the EFT to be used in a top-down way, to improve
the theoretical predictions in a specific BSM scenario.

\section{Acknowledgements}

We would like to thank Rikkert Frederix and Stefano Frixione for discussions
regarding the  parton shower treatment of resonances.  We are thankful to Paolo
Torrielli for help with technical tools.  We are also grateful to Peter Galler,
Sunghoon Jung, David Lopez-Val, and William Marciano for inspiring discussions.
We thank Fabio Maltoni for continuously supporting this work and for providing
valuable feedback.  D.B.F.~thanks Steffen Schumann for discussions about parton
shower limitations for the interference description. 
D.B.F.~acknowledge partial funding by the European Union as part of the H2020
Marie Sklodowska-Curie Initial Training Network MCnetITN3 (722104).
E.V.~is supported by the
research programme of the Foundation for Fundamental Research on Matter (FOM),
which is part of the Netherlands Organisation for Scientific Research (NWO).
C.Z.~is supported by the United States Department of Energy under Grant
Contracts DE-SC0012704, and by the 100-talent project of Chinese Academy of
Sciences.

\bibliography{bib}
\bibliographystyle{JHEP}

\end{document}